\documentclass[%
 aps,
 pra,
 reprint,
 amsmath,
 amssymb,
 groupedaddress,
 eprint,
 longbibliography,
 nofootinbib,
 nobibnotes,
]{revtex4-1}

\usepackage[utf8]{inputenc}
\usepackage{graphicx}
\usepackage{bm}
\usepackage{xcolor}
\usepackage{enumerate}
\usepackage{amsfonts, amsmath, amssymb}
\usepackage[english]{babel}
\usepackage{dsfont}
\usepackage{placeins}
\usepackage{comment}
\usepackage{mathrsfs}
\usepackage{enumerate}

\usepackage{hyperref}
\hypersetup{
	colorlinks = true,
	linkcolor = blue,
	citecolor = blue,
	urlcolor = blue,
	filecolor = black,
	linktoc = all,
}

\newcommand{\pp}{\partial}
\newcommand{\ph}{\varphi}

\newcommand{\HH}{\mathcal{H}}
\newcommand{\scrj}{\mathcal{J}}
\newcommand{\adag}{a^{\dagger}}
\newcommand{\dd}{\delta}
\newcommand{\ket}[1]{|#1\rangle}
\newcommand{\vecspan}{\textrm{span}}

\newcommand{\qm}{\textsc{qm}}

\newcommand{\bh}{\textrm{bh}}
\newcommand{\out}{\textrm{out}}
\newcommand{\inn}{\textrm{in}}
\newcommand{\rad}{\textrm{rad}}

\newcommand{\bra}[1]{\langle #1 |}
\newcommand{\MM}{\mathcal{M}}

\newcommand{\Sext}{\mathcal{S}_{\rm ext}}
\newcommand{\eh}{\textsc{eh}}

\begin{document}


\title{Unitarity and the information problem in an explicit model of black hole evaporation}

\author{Joseph Schindler}
\email{jcschind@ucsc.edu}

\author{Evan Frangipane}
\email{efrangip@ucsc.edu}

\author{Anthony Aguirre}
\email{anaguirr@ucsc.edu}
\affiliation{University of California Santa Cruz, Santa Cruz, CA, USA}

\date{\today}

\begin{abstract}
We consider the black hole information problem in an explicitly defined spacetime modelling black hole evaporation. Using this context we review basic aspects of the problem, with a particular effort to be unambiguous about subtle topics, for instance precisely what is meant by entropy in various circumstances. We then focus on questions of unitarity, and argue that commonly invoked semiclassical statements of long term, evaporation time, and Page time ``unitarity'' may all be violated even if physics is fundamentally unitary. This suggests that there is no horizon firewall. We discuss how the picture is modified for regular (nonsingular) evaporation models. We also compare our conclusions to recent holographic studies, and argue that they are mutually compatible.
\end{abstract}


\maketitle


\section{Introduction}

Whether information is preserved or lost during black hole evaporation has now remained an unresolved question for several decades~\cite{Hawking:1976ra,Hawking:1974sw,Page:1979tc,Hawking:1982dj,Zurek:1982zz,Carlitz:1986ng,Preskill:1992tc,Susskind:1993if,Bekenstein:1993bg,Page:1993df,Page:1993wv,Stephens:1993an,Strominger:1994ey,Polchinski:1995ta,tHooft:1996rdg,Horowitz:1996tx,Mikovic:1996bh,Mikovic:1997xm,Hajicek:1999dg,Giddings:2004ud,Horowitz:2003he,Hawking:2005kf,Russo:2005aw,Giddings:2006sj,Hayden:2007cs,Mathur:2009hf,Hossenfelder:2009xq,Mathur:2011uj,Almheiri:2012rt,Brustein:2012jn,Cai:2012um,Hossenfelder:2012mr,Page:2013dx,Good:2013lca,Bardeen:2014uaa,Haggard:2014rza,Harlow:2014yka,Bianchi:2014qua,Lochan:2015oba,Hawking:2016msc,Good:2016oey,Chakraborty:2017pmn,Marolf:2017jkr,Polchinski:2016hrw,Unruh:2017uaw,Bardeen:2018omt,Stoica:2018uli,Wallace:2017wzs,Amadei:2019wjp,Amadei:2019ssp,Rovelli:2019tbl,Penington:2019npb,Almheiri:2019hni,Almheiri:2020cfm,Almheiri:2019psf,Ashtekar:2020ifw,Bousso:2019ykv,Schindler:2019mub,Chen:2019uhq,Akers:2019nfi,Good:2019tnf,Chen:2020nag,Gan:2020nff,Kiefer:2020cbu,Maldacena:2020ady,Nomura:2019dlz,Gautason:2020tmk,Gaddam:2020rxb,Marolf:2020rpm,Krishnan:2020oun}. At the center of this issue is the principle of unitarity.

The precise statement of this principle depends on how a system is described. At the level of semiclassical gravity, there must be unitary evolution between the state of quantum fields on a family of Cauchy surfaces $\Sigma_u$ foliating some classical domain of dependence.%
\footnote{This form of unitarity holds by definition within a semiclassical theory, but is undefined when quantum gravity is not well described by a semiclassical approximation (\textit{e.g.}~due to correlations between matter and geometry). A semiclassical approximation is often used within standard discussions of black hole evaporation.}
At the level of quantum gravity, there must be unitary evolution between states in some underlying Hilbert space $\HH^\qm$ from which spacetime and gravity may emerge.
These statements are core principles of the quantum description of an information-preserving system, and we will assume them both to hold.

These core statements are not, however, the forms of ``unitarity'' usually invoked in relation to black hole information loss and the related firewall paradox. More commonly one asks:
\begin{itemize}
    \item Is there a unitary scattering matrix from past to future null infinity?
    \item Is the Hawking radiation in a pure state when evaporation ends?
    \item Does the entropy of Hawking radiation decrease at late times during evaporation?
\end{itemize}
We call these the questions of ``long term,'' ``evaporation time,'' and ``Page time'' unitarity, respectively; they will be made precise later.

Each of these questions is traditionally framed, by definition, in the context of semiclassical gravity---that is, in terms of quantum fields on a classical background. One reason they are difficult to resolve is that, since black hole physics involves strong quantum gravity effects, it is not clear what background spacetime (if any) can be used to model the process.

\begin{figure}[t]
    \centering
    \includegraphics[height=2.1in]{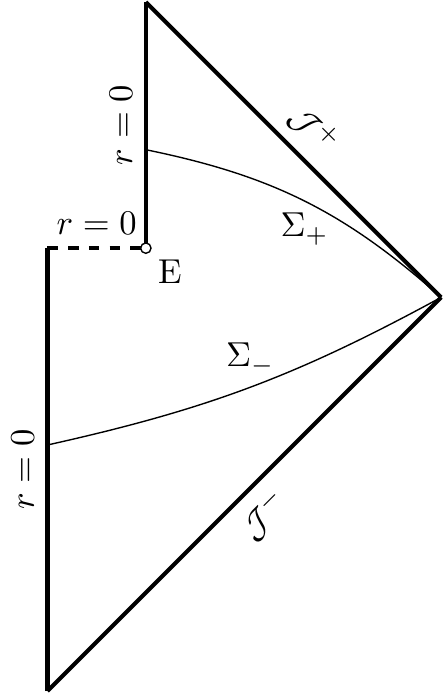}
    \caption{A Penrose diagram often associated with the process of black hole formation and evaporation.}
    \label{fig:basic}
\end{figure}

Lacking a known semiclassical solution, assumptions about a background spacetime generally come in the form of Fig.~\ref{fig:basic}. This diagram depicts a global causal structure and is useful for many purposes. Yet it is also problematic, in that it does not represent any particular physical model of the formation/evaporation process. For this reason it is difficult to make concrete statements about the geometry, and diagrams of this type are free to reflect biases of the artist.

In this article we will consider basic aspects of the black hole information problem, focusing on analyzing the various forms of ``unitarity,'' in the context of a particular background model for an evaporating black hole. While not an exact semiclassical solution, this model is at least a concretely defined metric that is likely similar to one (see Sec.~\ref{sec:classical-framework} for further discussion). Our hope is that framing these basic issues within a concretely defined context will help clarify their essential aspects, and provide a clearer grounds to confirm or refute assumptions and results about the evaporation process.

Much of the content of the article is review presented in a somewhat pedagogical manner. This is intentional. Our view is that subtle differences in understanding about foundational assumptions of the theory can propagate misunderstanding about the information issue, and therefore that we should clarify the framework being used. In this direction we make a particular effort to be clear about what Hilbert spaces define the quantum theory, what decompositions of the total Hilbert space are used when, and precisely what is meant by ``entropy of the black hole'' in various circumstances.

In the course of reviewing the basic issues in this context, we draw some conclusions that are not yet as widely accepted in the literature as we think they should be. We will argue that even assuming the fundamental notions of semiclassical and quantum gravitational unitarity do both hold, the more common notions of long term, evaporation time, and Page time unitarity can nonetheless fail. And while long term unitarity may be restored by appealing to regular (nonsingular) black hole models, the latter two cannot. These failures are sometimes said to represent ``information loss,'' but they in no way violate the underlying principles of unitarity.

An essential aspect of our arguments is the distinction between the semiclassical%
\footnote{By ``semiclassical Hilbert space'' throughout the paper, we really mean the Hilbert space of quantum field theory on the classical background $\MM$ (of Sec.~\ref{sec:classical-framework}), which resembles but is not a solution of semiclassical gravity. The word semiclassical in this context denotes the space of quantum fields on a background metric, distinguishing this space from a truly quantum gravitational one.}
Hilbert space $\HH(\Sigma_u)$ and an underlying quantum gravitational Hilbert space $\HH^\qm$. In particular we argue that the Page curve arises in the quantum gravitational Hilbert space, but not necessarily in the semiclassical one. In Sec.~\ref{sec:holographic} we consider further how these two levels of description are related, comparing our results to some recent results based on holography~\cite{Penington:2019npb,Almheiri:2019hni,Almheiri:2020cfm} and arguing that they are mutually compatible.

We are not the first to conclude that long term, evaporation time, and Page time ``unitarity'' can all be violated in a unitary theory, even if this conclusion has yet to be widely accepted in the literature.  And discussions akin to ours have appeared before in various places~\cite{Hossenfelder:2009xq,Unruh:2017uaw,Wallace:2017wzs,Stoica:2018uli,Rovelli:2019tbl,Gan:2020nff,Ashtekar:2020ifw}.  The primary novelty of this work is its meticulous framing and detailed presentation in a new, explicitly defined, background. Given the persistent controversy surrounding these topics, we hope this can be a useful step towards consensus.

\section{Quantum framework}
\label{sec:quantum-framework}

\subsection{Hilbert space of a partial Cauchy surface}
The Hilbert space of a quantum field theory in curved spacetime is generally defined (taking Birrell and Davies~\cite{Birrell:1982ix} as a canonical reference) as a Fock space of mode solutions to the free part of the classical field equation~\cite{Birrell:1982ix,DeWitt:1975ys,Gibbons:1978dw,Birrell:1978pk}. Here we will extend the standard formalism straightforwardly to consider partial (as opposed to global) Cauchy surfaces.

The basic approach is as follows. To each (orthonormal, positive frequency) complete set of classical modes $\xi_k$ on (the domain of dependence of) a hypersurface $\Sigma$ is associated a Fock space $\HH_{\xi}(\Sigma)$ on which the quantum field theory can be defined. Given any two sets of modes $\xi$, $\xi'$, complete on hypersurfaces $\Sigma$, $\Sigma'$ respectively, we say the Hilbert spaces 
\begin{equation}
\label{eqn:equiv}
    \HH_{\xi}(\Sigma) \sim \HH_{\xi'}(\Sigma')
\end{equation}
are physically equivalent whenever the domains of dependence $D(\Sigma) = D(\Sigma')$ are equal. The space 
\begin{equation}
\label{eqn:HSigma}
    \HH(\Sigma)
\end{equation}
is then defined as the equivalence class of all $\HH_\xi(\Sigma)$. One expects unitary transformations between semiclassical Hilbert spaces only when they are physically equivalent.

The requirement that a set of modes defining the Hilbert space be complete is essential. For instance, the set of outgoing Hawking modes is not complete on any relevant Cauchy surface, and must be embedded within a larger set of modes when analyzing the final state.

Some parts of this construction (e.g. continuous tensor product below) are only mathematically well-defined after a UV or IR cutoff is included. We assume such cutoffs can be applied where necessary.

\subsection{Hilbert space details}

The Hilbert space construction outlined above is a simple formalization of (sometimes implicitly used) standard methods. Nonetheless, we elaborate the details here for maximal clarity. For concreteness, consider the matter action
\begin{equation}
\label{eqn:matter-action}
    S= \tfrac{1}{2} \int d^4x \sqrt{|g|} \; \pp_\mu \phi \; \pp^\mu \phi
\end{equation}
defining a free real massless scalar field.

Consider a spatial hypersurface $\Sigma$ in spacetime, which is a Cauchy surface for its domain of dependence $D(\Sigma)$ (which may or may not be the entire spacetime). 

Let $\xi_k$ denote an orthonormal complete set of positive frequency modes%
\footnote{By ``a complete set of orthonormal positive frequency modes on~$\Sigma$'' we mean: a set $\xi_k$ of complex-valued solutions to the classical field equation such that $\varphi = \sum_k \big( c_k \, \xi_k + d_k \, \xi_k^* \big)$ is a classical solution matching arbitrary complex Cauchy data on~$\Sigma$, where $c_k$ and $d_k$ are complex coefficients, and such that $(\xi_{k'}, \xi_{k}) = \dd_{k'k}$ and $(\xi_{k'}, \xi_{k}^{*}) = 0$ in the inner product induced by the equations of motion (\textit{i.e.} the Klein-Gordon norm). The inner product is linear in the first argument and obeys $(\ph_1,\ph_2) = (\ph_2,\ph_1)^*=-(\ph_2^*,\ph_1^*)$. One way to find a suitable set of modes is to require $\omega > 0$ relative to some timelike Killing vector field, if one exists. More broadly these conditions have little to do with frequency, but rather relate the classical symplectic to the quantum structure. See~\cite{DeWitt:1975ys,Birrell:1982ix,Ashtekar:1975zn}.}
on $\Sigma$. To each mode are associated creation and annihilation operators $\adag_k$ and $a_k$ with canonical commutation relations $[a_{k}, \adag_{k}] = 1$. 

The Hilbert space $\HH_{\xi_k}$ of each mode is generated from a vacuum state $\ket{0_k}$ (defined by $a_k \ket{0_k}=0$) by the creation operators. Explicitly, for the bosonic field,%
\footnote{For a fermionic field one uses canonical anticommutation relations, resulting in $\HH_{\xi_k} = \vecspan \left( \{ \ket{0_k}, \ket{1_k}\} \right)$, a two-level system at each mode (in contrast to the bosonic case of a harmonic oscillator at each mode).}
\begin{equation}
    \HH_{\xi_k} = \vecspan \left( \{ \ket{n_k}, \, n_k \in \mathbb{N} \} \right)
\end{equation}
where \mbox{$\ket{n_k} \propto (\adag_k)^n \ket{0_k}$}. This basis obeys $N_k \ket{n_k} = n_k \ket{n_k}$ for the number operator $N_k =  \adag_k a_k$.

Denote by $\HH_{\xi}(\Sigma)$ the total Hilbert space of the modes $\xi$ on $\Sigma$, defined as the tensor product over modes
\begin{equation}
\label{eqn:Hmodes}
    \HH_{\xi}(\Sigma) = \otimes_k \; \HH_{\xi_k}.
\end{equation}
A basis for this space can be written $\ket{n_{k_1} n_{k_2} \ldots}$ (if the modes have a discrete index), which can also be translated to an equivalent Fock state notation. In the product space, the canonical commutation laws extend to $[\adag_{k'}, \adag_{k}] = [a_{k'}, a_{k}] = 0$ and $[a_{k'}, \adag_{k}] = \dd_{k'k}$. The quantum operator $\phi$ at each point then acts on this Hilbert space as
\begin{equation}
\label{eqn:phi}
    \phi = \sum_k \big( a_k \, \xi_k + \adag_k \, \xi_k^* \big),
\end{equation}
defining the quantum theory. 
Bogoliubov transformations derive from requiring (\ref{eqn:phi}) be equal in two sets of modes. The orthonormal positive frequency condition ensures commutators are preserved in the transformation.
The above statements translate to the case of a continuous index by standard methods~\cite{Birrell:1982ix}.

Consider now a partial Cauchy surface $\Sigma_{AB} = \Sigma_A \cup \Sigma_B$ that is the union of two disjoint subsurfaces. Choose a set of modes $\xi_{AB} = \xi_A \oplus \xi_B$ (here $\oplus$ is merely a suggestive notation for the union of two sets of modes) where $\xi_A$ is a set of modes on $\Sigma_{AB}$ with Cauchy data equal to zero everywhere on $\Sigma_B$ (and likewise for $\xi_B$). Given such a set of modes, it follows from (\ref{eqn:Hmodes}) immediately that
\begin{equation}
\label{eqn:HmodesAtimesB}
    \HH_{\xi_{A} \oplus \xi_{B}}(\Sigma_{AB}) = \HH_{\xi_{A}} \otimes \HH_{\xi_{B}}.
\end{equation}
But as $\xi_A$ alone forms a complete set of modes on~$\Sigma_A$, there is a natural identification of $\HH_{\xi_{A}}$ with $\HH_{\xi_{A}}(\Sigma_{A})$, and likewise for $B$. Thus we can write, in the sense of (\ref{eqn:equiv}--\ref{eqn:HSigma}), that
\begin{equation}
\label{eqn:HAtimesB}
    \HH(\Sigma_{AB}) = \HH(\Sigma_A) \otimes \HH(\Sigma_B).
\end{equation}
So even in the mode construction, Hilbert space may be built up as the tensor product of local subsystems.%
\footnote{One complete set of orthonormal positive-frequency modes on $\Sigma$ is given by a set $h_{x'}(x) = f_{x'}(x) + i g_{x'}(x)$, labelled by $x' \in \Sigma$, where $f_{x'}$ and $g_{x'}$ are classical solutions with $\delta$-function initial data at $x'$ in the field value and time-derivative respectively. (These are propagators of the homogeneous field equation, related to Wightman functions~\cite{Birrell:1982ix}.) Then for $x \in \Sigma$, in terms of these modes, $\phi(x) = a_x^\dag + a_x$. These modes give meaning to the expression $\HH(\Sigma) = \otimes_{x \in \Sigma} \HH_{x}$, a fully local decomposition of the Hilbert space. One can decompose $\HH(\Sigma_A \cup \Sigma_B) = \HH(\Sigma_A) \otimes \HH(\Sigma_B)$ in the same way---this is the construction usually implicitly or explicitly used in calculations of local von Neumann entropy. \label{foot:local-mode}
}

\section{(Semi-)Classical framework}
\label{sec:classical-framework}

Ideally one would study the information problem in a classical spacetime background that is an exact solution ($G_{\mu\nu} = 8 \pi \langle T_{\mu \nu} \rangle_{\rm ren}$) of semiclassical gravity with some quantum matter fields.%
\footnote{$\langle T_{\mu \nu} \rangle_{\rm ren}$ is a renormalized expectation value $\bra{\psi} T_{\mu \nu} \ket{\psi}$ of the stress tensor for the matter fields in whatever quantum state $\ket{\psi}$ the fields are in. $G_{\mu\nu}$ is the usual classical Einstein tensor.}
But due to the difficulty of incorporating the Hawking radiation backreaction, such a solution is not available.

Nonetheless, one can obtain approximations to such a solution using facts known from partial semiclassical calculations. For concreteness we continue to work with the massless scalar field (\ref{eqn:matter-action}), although the exact fields considered should not be essential to the picture.

We therefore define a spacetime $\MM$ (described below), which is one such approximation, to serve as a classical background for quantum fields in an approximately semiclassical context.

While the global causal structure of $\MM$ is the same as of Fig.~\ref{fig:basic}, the benefit of using an explicit model is that one may make definite statements about internal structure---in particular clarifying the status of the apparent and event horizons, the energy flux due to Hawking radiation, and the Schwarzschild mass at each spacetime point, within the model. These details allow one to construct a physically meaningful foliation in which to discuss the evaporation process, and provide additional intuition about how quantum calculations may be interpreted within the classical background. Equally importantly, the use of a concrete model precludes one from introducing potentially biased or self-contradictory assumptions about a background spacetime. So while $\MM$ is by no means assumed to exactly represent the spacetime structure of an evaporating black hole, it provides an explicit representation that is likely more useful than the vague model used implicitly in figures like Fig.~\ref{fig:basic}.

\subsection{The spacetime $\MM$}
\label{sec:setup:spacetime}
 To avoid a full technical treatment, we formally define $\MM$ as follows: let $\MM$ be the spacetime defined by Sec.~VI~of~\cite{Schindler:2019mub} but with $l=0$ (Schwarzschild interior). Presently we will give a more useful description.

\begin{figure}[t]
    \centering
    \includegraphics[scale=.73]{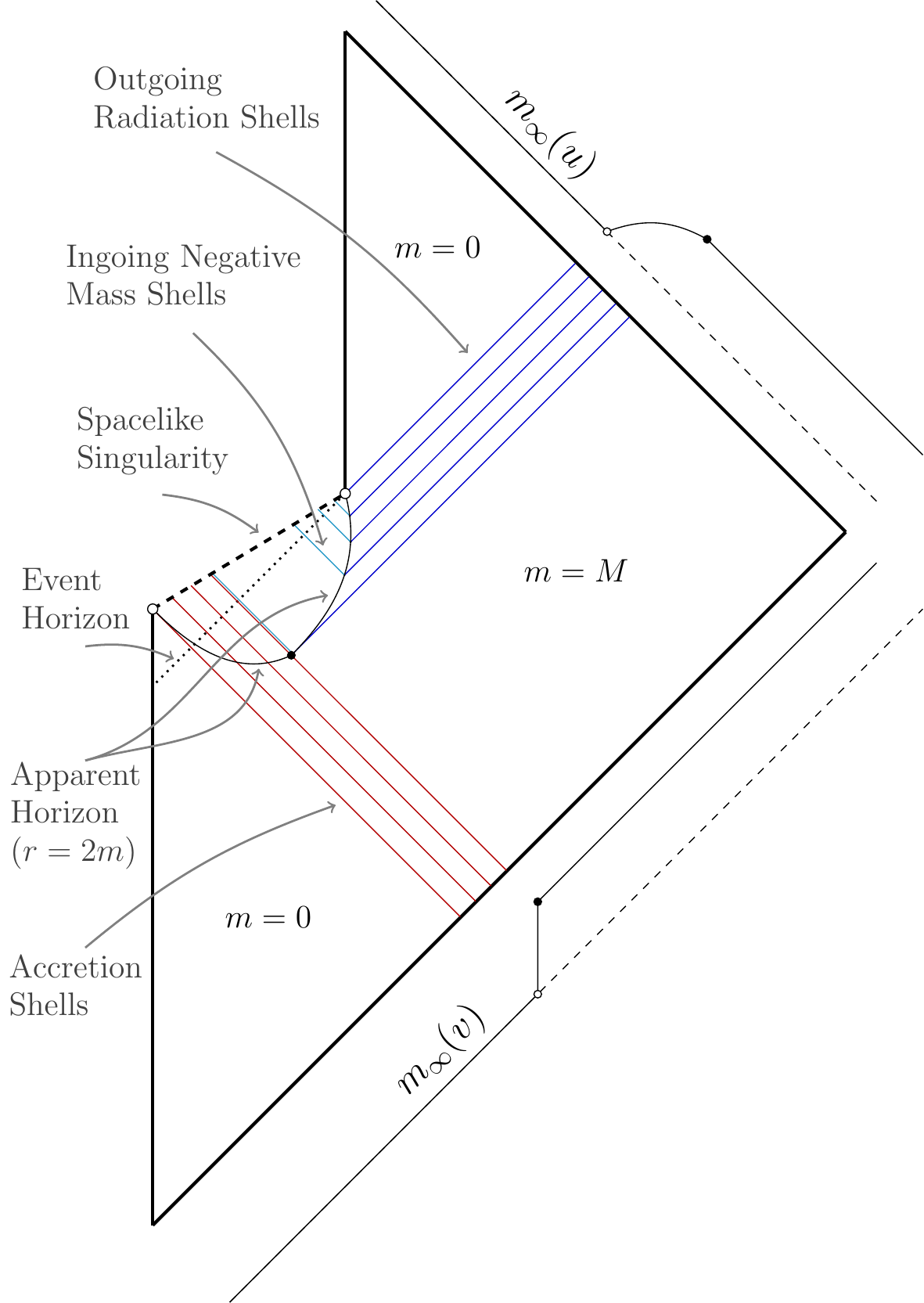}
    \caption{The spacetime $\MM$ used as a background for quantum fields. This schematic causal diagram for $\MM$ is known to be qualitatively correct based on Penrose diagrams that were computed numerically in previous work~\cite{Schindler:2018wbx,Schindler:2019mub}.}
    \label{fig:classical-background}
\end{figure}

The structure of $\MM$ is depicted in Fig.~\ref{fig:classical-background}. Locally the metric has the Schwarzschild form%
\footnote{The most conceptually simple representation of the metric is the local form (\ref{eqn:metric}), but there is no global coordinate system with this metric. For this reason we cannot simply write down $m(u,v)$ in a closed form. One can, however, parameterize future and past null infinity by continuous parameters that are locally $u,v$. The full metric is discussed in~\cite{Schindler:2019mub}. \label{foot:heuristically}}
\begin{equation}
\label{eqn:metric}
    ds^2 = - \left(1-\tfrac{2m(u,v)}{r(u,v)}\right) du \, dv + r(u,v)^2 \, d\Omega^2
\end{equation}
with a Schwarzschild mass $m(u,v)$ varying as a function of some null coordinates. The mass is a piecewise constant function forming an arbitrarily good stepwise approximation to some continuous dynamics. This leads to a shell ($\delta$-function) approximation to a smooth $G_{\mu\nu}$.

The mass function $m(u,v)$ is then chosen as follows (but see the caveat in Footnote~\ref{foot:heuristically}):
\begin{itemize}
    \item Formation occurs by collapse of a sequence of spherical shells, approximating a continuous accretion dynamics $m_{\infty}(v)$ as viewed from past null infinity. The total mass is $m_{\infty}(v\to \infty)=M$.
    
    \item Shells of outgoing radiation are emitted from (a~Planck length outside of)%
    \footnote{Self-consistency of the classical model does not allow emission to originate either inside of or further away from the apparent horizon at $r=2m$, see~\cite{Schindler:2019mub}. For this reason radiation must be emitted from near the apparent, and not the event, horizon.}
    the apparent horizon, approximating a continuous evaporation dynamics $m_{\infty}(u)$ as measured by an observer receiving the radiation at future null infinity. Corresponding shells of ingoing negative mass radiation fall in from (a Planck length outside of) the apparent horizon and are absorbed by the singularity. This model roughly approximates the DFU (Davies, Fulling, and Unruh~\cite{Davies:1976ei}) stress tensor for black hole evaporation.
    
    \item The Schwarzschild mass changes across shells as dictated by standard junction conditions and the DTR (Dray--'t Hooft--Redmount~\cite{Dray:1985yt,1985PThPh..73.1401R,Barrabes:1991ng}) relation.
\end{itemize}
This is in essence a discretized version of the model studied first by Hiscock~\cite{Hiscock:1981xb}. 

Note two subtle points about this spacetime. First, it is the apparent horizon, and not the event horizon, which lies at $r=2m$. The apparent horizon is spacelike during accretion and timelike during evaporation. Second, shells emitted from the horizon arise from an approximation to the DFU~\cite{Davies:1976ei} stress tensor. They are not chosen to directly model Hawking pairs. In particular, the Hawking modes behind the event horizon propagate parallel to it (as illustrated later), while the negative energy flux modelled by shells is directed transversely into the horizons.

\subsection{Globally hyperbolic subdomains}

\begin{figure}[t]
    \centering
    \includegraphics[height=2.5in]{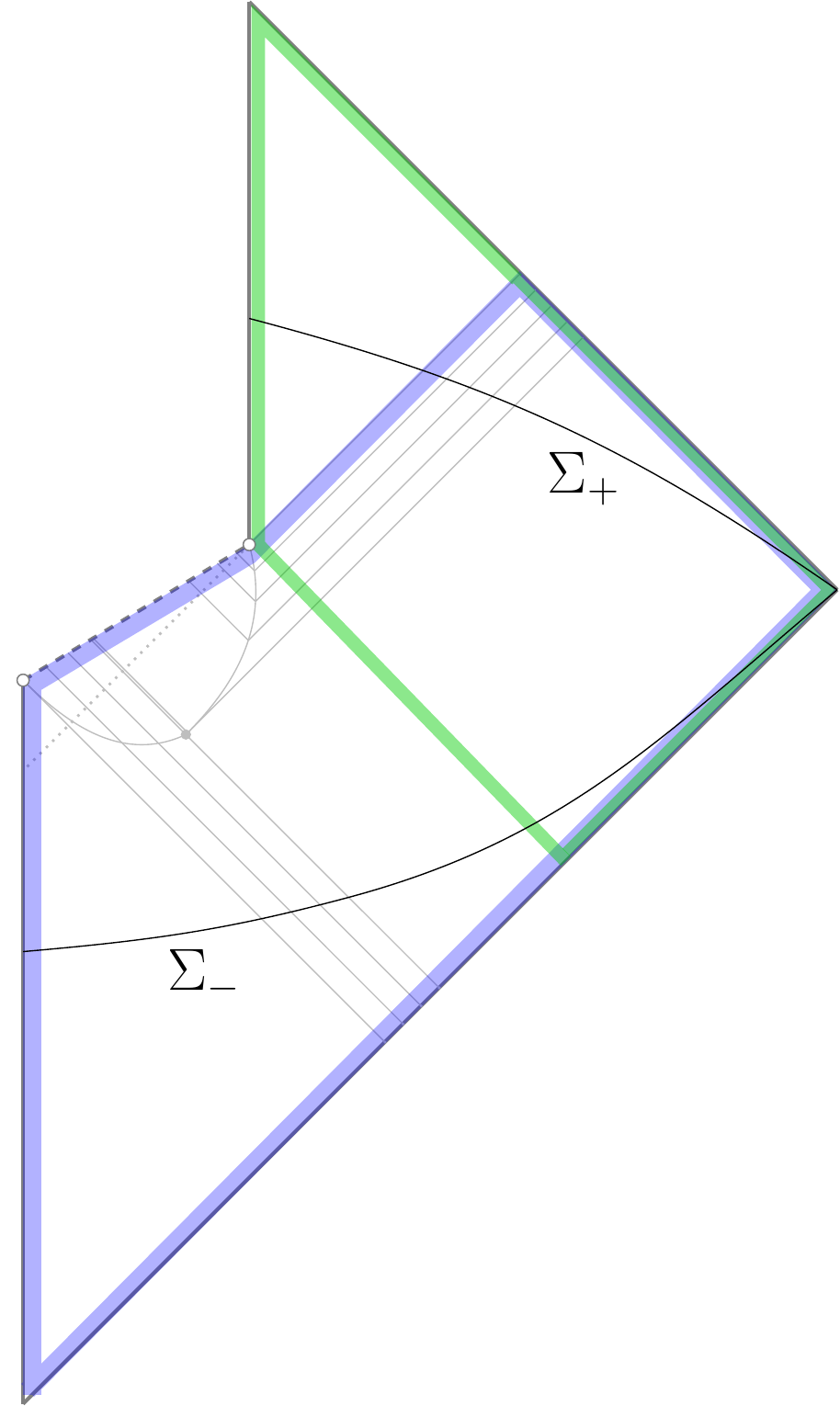}
    \caption{Early ($\Sigma_-$) and late ($\Sigma_+$) partial Cauchy surfaces in $\MM$. The domains of dependence $D(\Sigma_-)$ (blue outline) and $D(\Sigma_+)$ (green outline) are not equal.}
    \label{fig:long-term}
\end{figure}

$\MM$ is not globally hyperbolic.%
\footnote{See~\cite{Schindler:2019mub} for discussion of why no spacetime with the general structure of Fig.~\ref{fig:basic} is globally hyperbolic.}
As depicted in Fig.~\ref{fig:long-term}, early and late spatial slices have unequal domains of dependence.

An initial state describing collapsing matter on an early slice like $\Sigma_-$ can be propagated throughout the domain of dependence $D(\Sigma_-)$. Since this region contains the entire process of black hole formation and evaporation (up to the final moments), it is sufficient to focus on this region for much of the discussion of information loss. In particular the evaporation time and Page time unitarity questions depend only on this region.

\subsection{Foliation $\Sigma_u$ of $D(\Sigma_-)$}

\begin{figure}[t]
    \centering
    \includegraphics[height=2.5in]{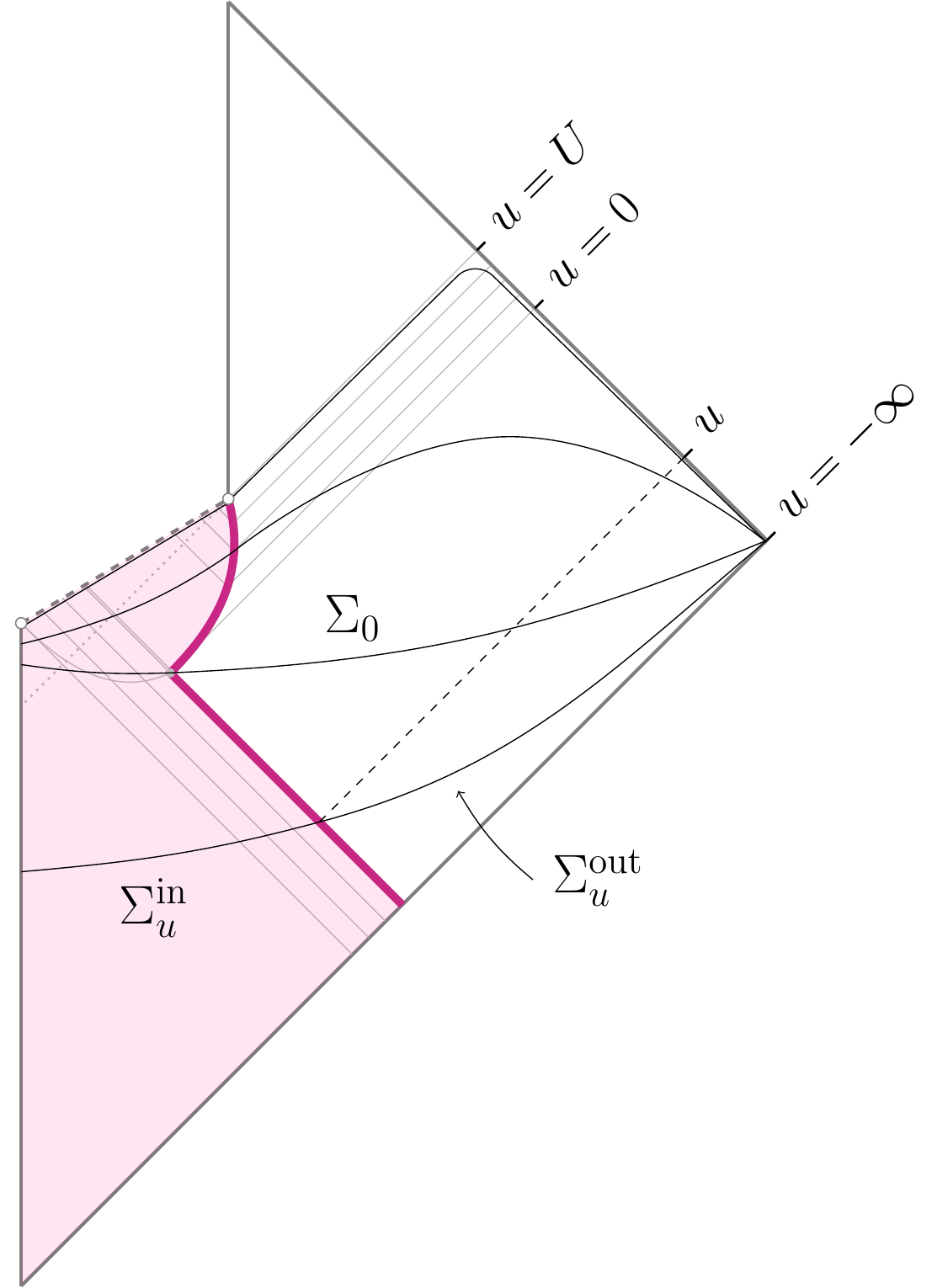}
    \caption{Foliation $\Sigma_u$ of the region $D(\Sigma_-)$ in $\MM$. The surface $\Sext$ (magenta), which coincides with the outermost accreting shell for $u \leq 0$ and with the outermost apparent horizon for $u \geq 0$ (and with both at $u=0$ where they meet), separates $D(\Sigma_-)$ into ``in'' (magenta fill) and ``out'' regions. Each \mbox{$\Sigma_u=\Sigma_u^\inn \cup \Sigma_u^\out$} decomposes accordingly. The label $u$ is the time at future null infinity (extrapolated along radial null curves) when each $\Sigma_u$ intersects $\Sext$. Evaporation begins at $u=0$ and ends at $u=U$.}
    \label{fig:foliated}
\end{figure}

Consider the domain of dependence $D(\Sigma_{-})$ in Fig.~\ref{fig:long-term}. This region is globally hyperbolic, and therefore can be foliated by a family of surfaces $\Sigma_u$ (each a Cauchy surface for $D(\Sigma_{-})$) as depicted in Fig.~\ref{fig:foliated}. This region contains the entire process of formation and evaporation, including all the Hawking radiation.

A surface $\Sext$ (the ``exterior surface of the collapsing matter/black hole'') separates $\MM$ into ``in'' and ``out'' regions. This surface is defined to coincide with the outermost accreting shell until it intersects the apparent horizon, after which it coincides with the outer part of the apparent horizon. Each
\begin{equation}
    \Sigma_u = \Sigma_{u}^{\inn} \cup \Sigma_{u}^{\out}
\end{equation}
decomposes into ``in'' and ``out'' surfaces accordingly.

Each $\Sigma_u$ is labelled by the time $u$ at future null infinity when it intersects $\Sext$. (One extrapolates the intersection to infinity along radial null curves.) By convention let $u=0$ denote the time when the outermost shell crosses the apparent horizon. The end of evaporation occurs some finite time $U$ later. Thus $u \in (-\infty, U)$ foliates the entire domain. Let $\Sigma_U$ denote some $\Sigma_u$ arbitrarily close to $u=U$.

\subsection{Hilbert space, modes, and states}
\label{sec:classical-framework:modes}

As in Sec.~\ref{sec:quantum-framework}, Hilbert spaces are defined as Fock spaces of classical modes. As is standard (see \textit{e.g.}~\cite{Hawking:1976ra,Hawking:1974sw,Good:2013lca}), we work in terms of modes describing classical wavepackets. Each set of modes $\xi$ we define is implicitly taken to represent a set $\xi_{i j l m}$ describing wavepackets centered at time $t_i$ and frequency~$f_j$ with temporal width scale $\sigma$ (times and frequencies being relative to a relevant coordinate system), and with angular harmonic component $Y_{lm}$. To obtain standard pair creation of Hawking modes requires appropriately coordinating wavepacket spectra across sets of modes. This can be done for static black holes~\cite{Hawking:1974sw,Hawking:1976ra} and we assume something analogous can be done here.

\begin{figure}[t]
    \centering
    \includegraphics[height=2.5in]{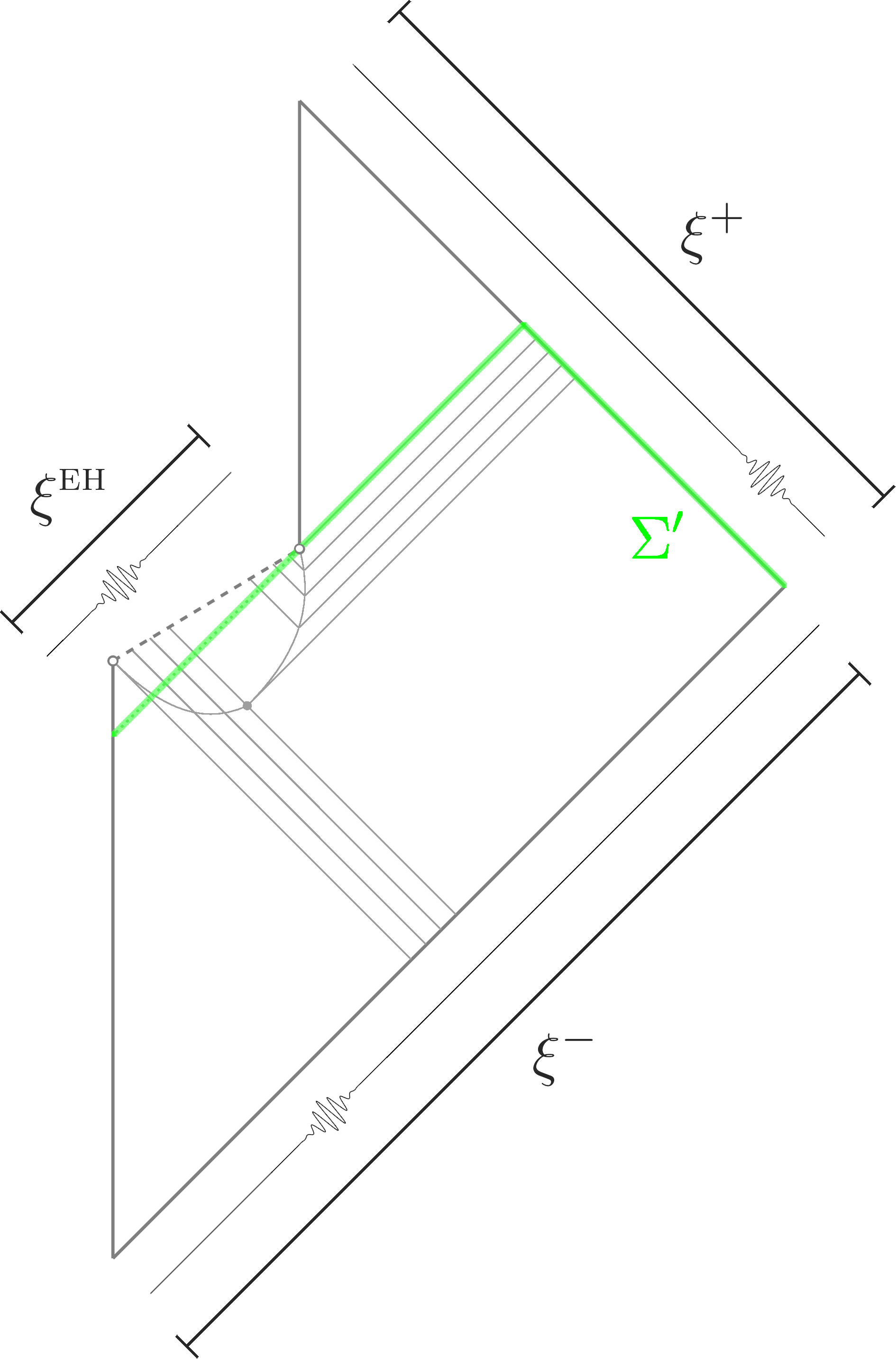}
    \caption{Sets of modes can be defined on various partial Cauchy surfaces in $\MM$. Modes on each surface are taken to represent classical wavepackets, as can be achieved by appropriate transformations from oscillating modes. Modes $\xi^-$ at past null infinity define the quantum state $\ket{\psi}$. Transforming to another set $\xi^\eh \oplus \xi^+$ dictates particle creation. Modes defined on $\Sigma'$ (green) can help resolve a pathology of the modes $\xi^\eh \oplus \xi^+$, which are technically not a valid complete set in $\MM$.}
    \label{fig:modes}
\end{figure}

Several relevant sets of modes are depicted in Fig.~\ref{fig:modes}.

The modes $\xi^-$ are orthonormal positive frequency ingoing wavepackets with respect to asymptotically flat coordinates at past null infinity. These provide a complete set on all $\Sigma_u$ and define $\HH(\Sigma_u)$.

The relevant quantum state%
\footnote{We are in the Heisenberg picture where the state $\ket{\psi}$ is fixed. Time dependence arises both in operators, and when the state is described in terms of a time-dependent mode decomposition.}
on $\HH(\Sigma_u)$ is usually taken to be an initial vacuum state $\ket{0_-} \equiv \ket{0}_{\xi^-}$. However we can just as well allow for the more general state (splitting $\xi^-$ into sets of wavepackets before, during, and after the presence of collapsing shells)
\begin{equation}
\label{eqn:psi}
    \begin{array}{rcllllll}
        \ket{\psi} &=& &
        \ket{0}_{\xi^{-}_{\rm before}} 
        &\otimes&
        \ket{\psi}_{\xi^{-}_{\rm collapse}} 
        &\otimes&
        \ket{0}_{\xi^{-}_{\rm after}}
    \end{array}
\end{equation}
to include a description of the collapsing matter. Particle creation in excited states such as this is closely related to that in vacuum~\cite{Carlitz:1986nh}.

The modes $\xi^+$ and $\xi^\eh$ (Fig.~\ref{fig:modes}) will also be relevant. $\xi^+$~are purely outgoing positive frequency wavepackets relative to asymptotically flat coordinates at future null infinity, with zero Cauchy data at the event horizon. $\xi^\eh$~have purely ingoing Cauchy data at the event horizon, with zero Cauchy data at future null infinity. $\xi^\eh$ modes can be formed into ``wavepackets'' with a particular correspondence to those in $\xi^+$ (at least in the quasistatic approximation~\cite{Hawking:1976ra}).

To analyze particle creation by the metric, one performs a Bogoliubov transformation from the modes $\xi^-$ defining $\ket{\psi}$ to some other complete set of orthonormal positive-frequency modes. Technically $EH \cup \scrj^+$ is not a Cauchy surface for $D(\Sigma_u)$ due to causal curves propagating from $\Sigma_u$ to the singular point at the endpoint of evaporation. But (as discussed in Sec.~\ref{sec:semiclassical:pathology}), as is commonly done, let us ignore this technicality and assert that $\xi^\eh \oplus \xi^+$ forms a complete set of modes on~$\Sigma_u$ that can be used for this purpose.

In analogy with the standard Hawking calculation~ \mbox{\cite{Hawking:1974sw,Hawking:1976ra}}, one expects that in terms of $\xi^\eh \oplus \xi^+$, the state $\ket{\psi}$ contains entangled pairs of ingoing (in $\xi^\eh$) and outgoing (in $\xi^+$) Hawking modes.

\begin{figure}[t]
    \centering
    \includegraphics[height=2.5in]{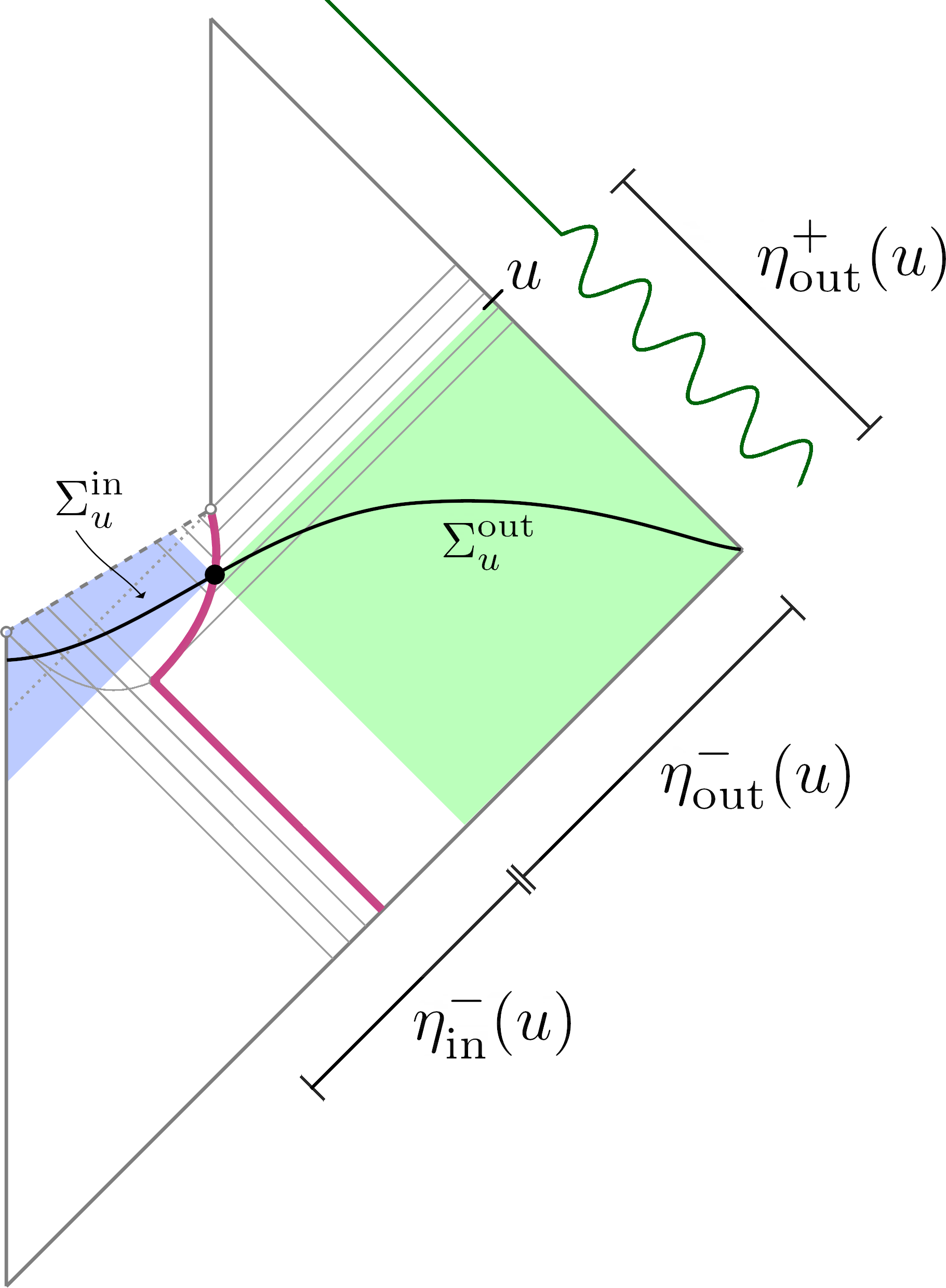}
    \caption{Splitting $\HH(\Sigma_u)$ into time-dependent \textit{in} and \textit{out} subspaces. The $\eta$ modes are wavepackets with limited support (support illustrated by green curve). Modes $\eta^\inn_u$ and $\eta^\out_u$ (see~(\ref{eqn:modes-in-out})) are respectively complete on the \textit{in} (blue fill) and \textit{out} (green fill) domains of dependence. All these spaces are time-dependent, with the boundaries and mode support regions sliding around as functions of $u$. As $u \to U$ the \textit{in} domain becomes the region behind the event horizon.}
    \label{fig:in-out-modes}
\end{figure}

\subsection{``In'' and ``Out'' Hilbert spaces}
\label{sec:classical-framework:inout}

To define a decomposition 
\begin{equation}
\label{eqn:hilbert-in-out}
    \HH(\Sigma_u) = \HH(\Sigma_u^\inn) \otimes \HH(\Sigma_u^\out)
\end{equation} 
into time-dependent \textit{in} and \textit{out} Hilbert spaces requires a set of modes $\eta_u = \eta^\inn_u \oplus \eta^\out_u$, where each mode only has support in the relevant subregion (see Sec.~\ref{sec:quantum-framework}).

The usual way to construct this set is from modes with Cauchy data localized at each point (see Footnote~\ref{foot:local-mode}). However the same can be achieved using wavepackets from infinity, cut off to have support only in the relevant region. This method makes the local and global Hilbert space constructions more similar.

Thus define modes $\eta_u$ as follows (see Fig.~\ref{fig:in-out-modes}). From complete sets of oscillating modes with limited support (the support is a function of $u$ as illustrated in the figure) at past and future null infinity, construct sets of orthonormal positive frequency wavepackets $\eta_\inn^-(u)$, $\eta_\out^-(u)$, and $\eta_\out^+(u)$. Then
\begin{equation}
\label{eqn:modes-in-out}
\begin{array}{l}
    \eta^\inn_u \equiv \eta_\inn^-(u) ,
    \\[6pt]
    \eta^\out_u \equiv \eta_\out^-(u) \oplus \eta_\out^+(u),
\end{array}
\end{equation}
are complete on $\Sigma^\inn_u$, $\Sigma^\out_u$ respectively (\textit{cf.} Fig.~\ref{fig:in-out-modes} and Eq.~(\ref{eqn:HAtimesB})). This suffices to define (\ref{eqn:hilbert-in-out}). 

The $\eta$ modes can be taken to be similar to the $\xi$ modes except near the boundaries where support is cut off, where $\eta$ modes are non-analytic.

\subsection{Apparent vs. event horizon}

One may be tempted to use the event horizon, rather than the apparent horizon, to define the horizon area and \textit{in} and \textit{out} regions. In addition to the fact that the apparent horizon (at $r=2m$) has local properties while the event horizon is global, there are a few reasons not to do so.

First, as the event horizon lies entirely at $u=U$, there is no meaningful way to relate times at future infinity (\textit{i.e.}~for an observer receiving Hawking radiation) to areas on the event horizon. And second, if $\Sigma^\out_u$ were defined by the event horizon, the ``out'' region would contain all the Hawking radiation at all times. We will not consider this possibility further.

\subsection{Pathology at future null infinity}
\label{sec:semiclassical:pathology}

We have taken the Hawking modes to be a subset of the complete set $\xi^\eh \oplus \xi^+$ (Fig.~\ref{fig:modes}), implicitly assuming that $EH \cup \scrj^+$ is a Cauchy surface for $D(\Sigma_u)$ in $\MM$. This choice of modes is motivated by the requirement that late time modes be regular for observers at $\scrj^+$ (future null infinity). Its validity is usually justified by analogy with the standard Hawking calculation (for a static black hole formed by collapse) where $EH \cup \scrj^+$ is a global Cauchy surface.

We must emphasize, however, that $EH \cup \scrj^+$ is \textit{not} a Cauchy surface for $D(\Sigma_u)$ in $\MM$ (nor for $\MM$ globally), due to curves terminating at the open singular point at the end of evaporation. Therefore $\xi^\eh \oplus \xi^+$ is technically not a valid complete set of modes on $D(\Sigma_u)$. Despite this pathology, we proceed as if it were valid, in order to connect to existing parts of the literature. Modified versions of the arguments below can be made to apply to a more correct mode decomposition, but we will not do so here.%
\footnote{It would be more correct to regard Hawking modes in $\MM$ as part of a complete set defined on $\Sigma'$ in Fig.~\ref{fig:modes}. Technically $\Sigma'$ is also not a Cauchy surface for $D(\Sigma_u)$, but as the limit of a set of Cauchy surfaces it is admissible. Modes on $\Sigma'$ are genuinely different from $\xi^\eh \oplus \xi^+$, since each mode will have support in only a limited subset of $\scrj^+$, and thus be nonanalytic on $\scrj^+$ as a whole.}

Not only is $EH  \cup \scrj^+$ not technically a Cauchy surface, it fails badly at being one. If one tries to ``fill in'' the open singular point at the endpoint of evaporation (\textit{e.g.} by regularizing the singularity), the surface fails to remain achronal. If one tries to deform it to avoid the singular point, the same occurs.  And it is not the limit of any set of rigorous Cauchy surfaces. This pathology may be more than a benign technicality; for instance it shows that $\MM$ (and likely all its close relatives) is a counterexample to the ``PS Assumption'' of~\cite{Marolf:2020rpm}.

In Sec.~\ref{sec:long-term} we will discuss how ``long term unitarity'' violation is inherent to the spacetime structure of $\MM$. The pathology at future null infinity discussed above is another, more subtle, manifestation of the same effect. In order to obtain a non-pathological $\scrj^+$, one can regularize the singularity as in Fig.~\ref{fig:regular}. Then $\xi^+$ alone are a complete set of modes. In that case $\xi^\eh \oplus \xi^+$ double-counts event horizon modes, as $EH  \cup \scrj^+$ is not achronal.

\subsection{Unitarity questions}

The principle of unitarity in a semiclassical context implies unitary evolution between the state of quantum fields on set $\Sigma_u$ of Cauchy surfaces.%
\footnote{Given the Hilbert space construction above, this notion is almost trivial: the (Heisenberg) state $\ket{\psi}$ is fixed, while a choice of modes $\xi$ complete on $\Sigma_u$, used to define the Hilbert space basis, may vary with time. Unitarity then merely states that a unitary transformation relates valid complete bases. If one transforms to a Schrodinger wavefunctional picture (say through the local modes of Footnote~\ref{foot:local-mode}), this reduces to a standard statement of unitary evolution of states. This also ensures Heisenberg operators like (\ref{eqn:phi}) evolve unitarily in a time dependent mode basis.}
This principle holds absolutely within the present semiclassical framework.

However, several different forms of ``unitarity,'' arising on different time scales, are often considered relevant to discussions of black hole information loss. Assuming an initial pure state $\ket{\psi}$ on $\Sigma_{-}$  (Fig.~\ref{fig:long-term}), we say that evaporation is
\begin{itemize}
    \item \emph{Long term unitary} if there is a pure state on surfaces like $\Sigma_+$ (Fig.~\ref{fig:long-term}).
    \item \emph{Evaporation time unitary} if Hawking radiation is in a pure state at the end of evaporation.
    \item \emph{Page time unitary} if the entropy of Hawking radiation is decreasing with $S \leq A/4$ at late times.
\end{itemize}
In the following two sections we make these ideas mathematically precise, considering each in turn, and argue that none of them is expected to hold in $\MM$ at the semiclassical level. Moreover, regularizing the singularity in $\MM$ can restore the possiblility of long term unitarity, but not evaporation time or Page time unitarity.

One can frame these statements in terms of either the Hilbert space of Hawking modes, or in terms the Hilbert space of the \textit{out} region. We first consider the former, then return to the latter in Sec.~\ref{sec:in-out}.

\section{Long term unitarity}
\label{sec:long-term}

The ``long term'' unitarity question is the following: Is~there necessarily a unitary evolution from quantum states on $\Sigma_-$ to quantum states on $\Sigma_+$ in Fig.~\ref{fig:long-term}?%
\footnote{This deals with the state of semiclassical matter fields. A separate question is whether evaporation can be described by a unitary $S$ matrix in quantum gravity, \textit{e.g.}~in a path integral approach. These are not equivalent, in part because correlations may arise between the matter and geometry, but also because one might sum over geometries where the initial and final surfaces have different domains of dependence.}

\subsection{In $\MM$}

If physics is accurately described by semiclassical gravity on a background spacetime like $\MM$, the answer is clear: \emph{there is no reason to expect long term unitarity}. The domains of dependence $D(\Sigma_-) \neq D(\Sigma_+)$ are unequal, and therefore, as discussed in Sec.~\ref{sec:quantum-framework}, the Hilbert spaces 
\begin{equation}
    \HH(\Sigma_-) \neq \HH(\Sigma_+)
\end{equation}
are physically  inequivalent. Unitary evolution is not expected between physically inequivalent Hilbert spaces. Likewise, one does not expect invertible evolution of classical fields between surfaces with unequal domains of dependence.

It must be emphasized that there is no guarantee that black hole evaporation is accurately described by semiclassical gravity on a background spacetime like $\MM$. However, if one gives up the assumption that something like $\MM$ is correct---for example by demanding $\Sigma_-$ and $\Sigma_+$ have a unitary relation---one must also give up on using the spacetime diagram for $\MM$ to analyze the problem. (Or at least, provide some other justification for using such a diagram.) On occasion in the literature, studies will implicitly argue that the semiclassical description is  incomplete or incorrect, yet at the same time continue making essential use of diagrams based on semiclassical spacetimes like $\MM$ or Fig.~\ref{fig:basic}. The self-consistency of such arguments must be called into question.

\subsection{With a regularized singularity}
\label{sec:long-term:regular}

Models like $\MM$ have long term unitarity violation ``baked into'' their structure. One way to circumvent this issue is to replace $\MM$ with a globally hyperbolic spacetime obtained by regularizing the $r=0$ singularity. 

\begin{figure}[t]
    \centering
    \includegraphics[height=2.5in]{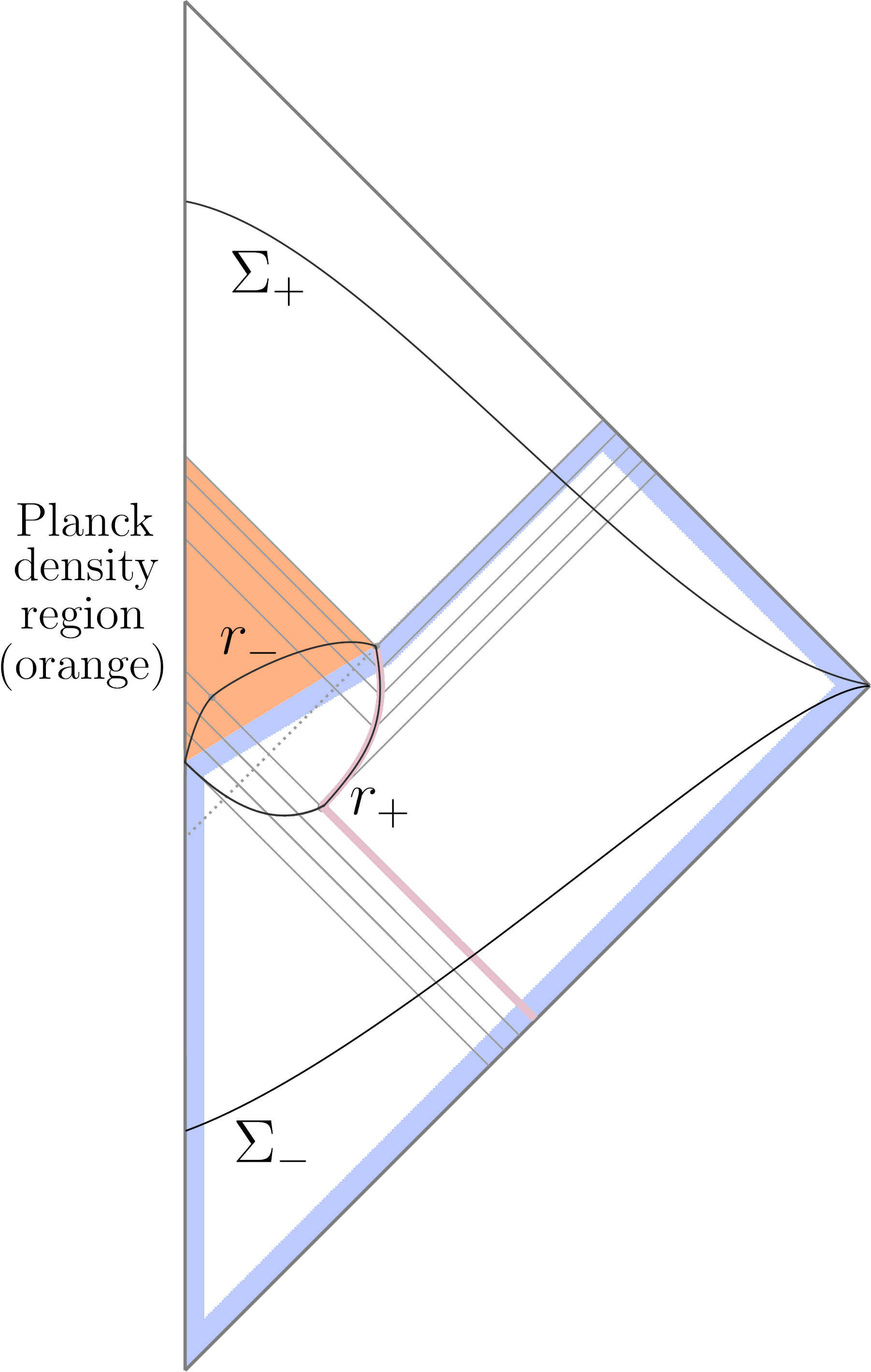}
    \caption{A spacetime like $\MM$ but with a regular (nonsingular) center (see~\cite{Schindler:2019mub} for a detailed exposition with numerically computed causal diagrams). An inner apparent horizon ($r_-$) lies within a Planck density core (orange). The evaporation time and Page time unitarity problems, which are described entirely within the foliation $\Sigma_u$ covering the early region (blue outline), are exactly the same here as in the singular case. Long term unitarity is viable in this background, unlike in~$\MM$, but depends on how initial data propagates through the strong quantum gravity region.}
    \label{fig:regular}
\end{figure}

One example of a regularized nonsingular background (based on a Hayward model~\cite{Hayward:2005gi,Schindler:2019mub}) is depicted in Fig.~\ref{fig:regular}. Models of this type are useful in that they include---rather than relegating to a singularity---a region of extreme density/curvature where quantum gravitational effects are important and known physics may fail.%
\footnote{Regular models also introduce other issues associated with the inner horizon and exposed core~\cite{Schindler:2019mub}. Note however that the future surface of the dense region, which appears large due to conformal transformations in the diagram, is actually Planckian in size---not so different from the naked singularity in $\MM$---and that the inner horizon is hidden within the dense quantum gravity region.}
For the same reason, semiclassical statements about such models must be taken with a grain of salt. The quantum gravity region may be thought of as a core that sources the gravitational field after collapse has completed.

The regularized model Fig.~\ref{fig:regular} does predict, at the semiclassical level, that long term unitary holds. The mechanism is uncertain, however, as semiclassical initial data would propagate through the quantum gravity region.

Unlike the long term issue, the evaporation time and Page time unitarity questions are framed entirely within the foliation $\Sigma_u$ of the early region (blue outline) in Fig.~\ref{fig:regular}. In this region the geometry is effectively identical to the singular case~\cite{Schindler:2019mub}. The discussion of evaporation time and Page time issues is therefore unaffected by regularizing the singularity. However entropy at infinity may then be purified after evaporation ends in such models.

\section{Evaporation time and Page time unitarity}
\label{sec:page-time}

This section discusses the ``evaporation time'' unitarity issue, which relates to the von Neumann entropy of Hawking modes at the end of evaporation, and the ``Page time'' information issue, which tracks the time dependence of this entropy throughout the evaporation process.

Arguments for a firewall usually assume that unitarity implies the von Neumann entropy of Hawking modes must follow a Page curve. We will argue that this is not the case: in the manifestly unitary semiclassical theory of fields on $\MM$, one should \textit{not} expect a Page curve for the entropy of Hawking modes.

There is an important distinction here: a Page curve \textit{should} be expected to arise in quantum gravitational descriptions---in particular, we do not disagree with recent holographic derivations~\cite{Penington:2019npb,Almheiri:2019hni,Almheiri:2020cfm} of the Page curve. But a Page curve in the underlying quantum gravity theory does not imply a Page curve for semiclassical Hawking modes---and it is the entropy of semiclassical modes whose Page curve implies a firewall. The connection to quantum gravity is explored further in Sec.~\ref{sec:holographic}.

\begin{figure}
    \centering
    \includegraphics[height=2.5in]{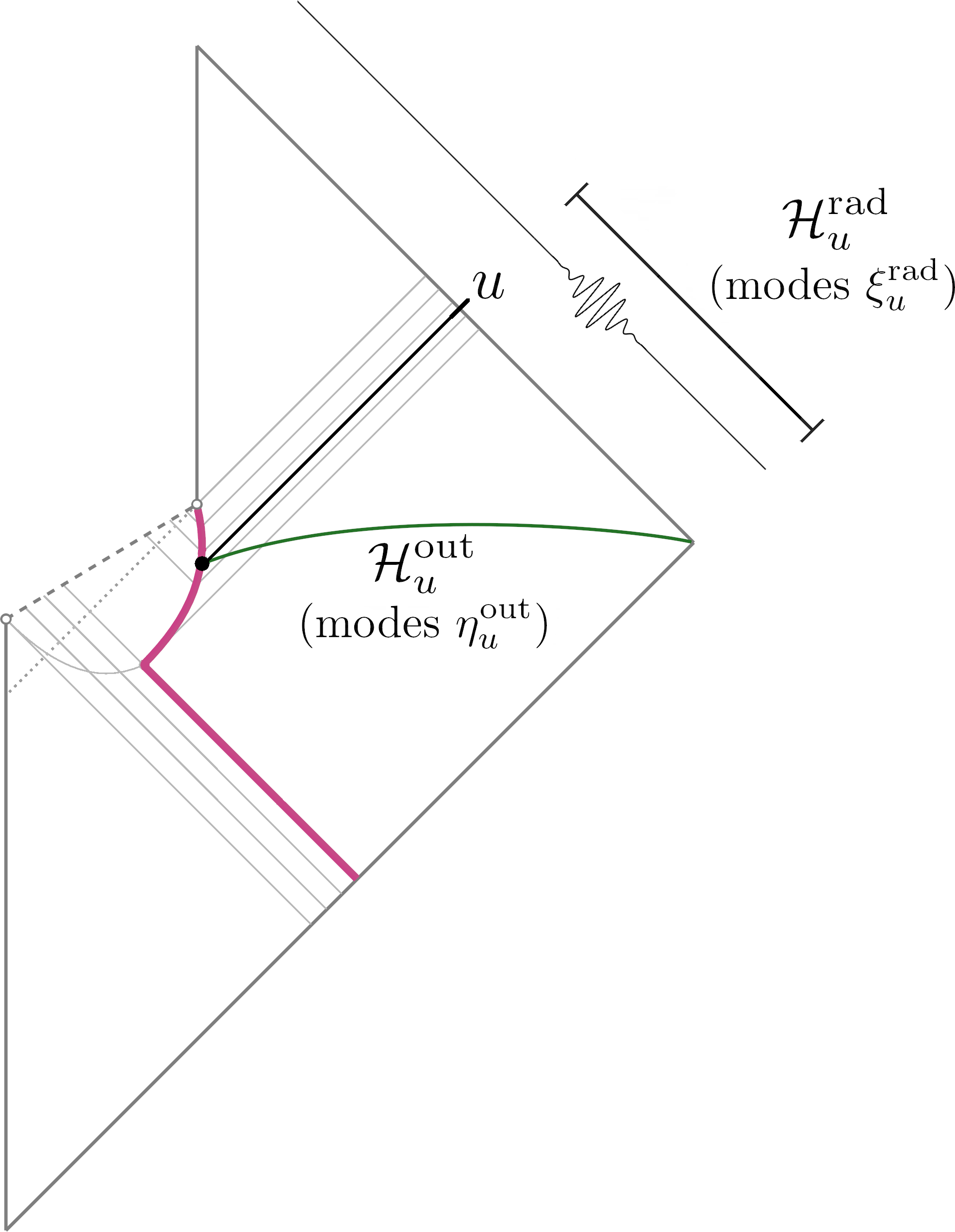}
    \caption{The entropy $S_\rad(u)$ of outgoing Hawking modes up to time~$u$ is defined as von Neumann entropy in $\HH^\rad_u$, which is the Hilbert space of the modes $\xi^\rad_u$. These $\xi^\rad_u$ are the subset of $\xi^+$ (Fig.~\ref{fig:modes}) with wavepackets centered before time $u$. A different entropy, the von Neumann entropy in the out region $\HH^\out_u \equiv \HH(\Sigma^{\out}_u)$, is discussed later.
    }
    \label{fig:basic-infinity-modes}
\end{figure}

\subsection{Entropy of Hawking modes}

Of interest here is the von Neumann entropy of Hawking modes as a function of time. We denote this entropy $S_\rad(u)$, a function of time $u$ at future null infinity, and define it as follows.

The modes labelled $\xi^\rad_u$ in Fig.~\ref{fig:basic-infinity-modes} are the ``Hawking modes up to time $u$.'' These are a subset of the wavepacket modes $\xi^+$ in Fig.~\ref{fig:modes}, specifically, the subset with wavepackets centered before time $u$. The ``Hilbert space of Hawking radiation at time $u$'' is the Hilbert space of these modes,
\begin{equation}
    \HH^\rad_u \equiv \HH_{\xi^\rad_u}.
\end{equation}
This Hilbert space can be written as the tensor product of Hilbert spaces describing wavepacket modes, defined at each time $u'$, over times $u'<u$. Note that $\HH^\rad_u$ is distinct from the Hilbert space of the \textit{out} region $\HH(\Sigma^\out_u)$, which will be discussed in Sec.~\ref{sec:in-out}.

The modes $\xi^\rad_u$ are a subset of $\xi^+$ and therefore of the full set $\xi^\eh \oplus \xi^+$ (\textit{cf.}~Sec.~\ref{sec:classical-framework:modes}). In this way the Hawking radiation Hilbert space $\HH^\rad_u$ is a subspace of the full semiclassical Hilbert space $\HH(\Sigma_u)$. Since the global state $\ket{\psi}$ in $\HH(\Sigma_u)$ is pure, the reduced state in $\HH^\rad_u$ will generically be mixed, with density matrix $\rho^\rad_u$.

The entropy of Hawking modes at time $u$ is then
\begin{equation}
    S_\rad(u) \equiv S(\rho^\rad_u),
\end{equation}
the von Neumann entropy in the Hawking radiation subspace of the semiclassical Hilbert space of fields.

\subsection{Evaporation time unitarity}

``Evaporation time unitary'' holds if
\begin{equation}
    S_\rad(U) = 0,
\end{equation}
in other words, if the Hawking modes are in a pure state at the time $U$ when evaporation completes. This form of unitarity is assumed in, \textit{e.g.}, the well known ``AMPS'' firewall paper~\cite{Almheiri:2012rt}.

Taken as an assumption in its own right, this would simply not be a correct application of the general principle of unitarity to semiclassical fields in the spacetime~$\MM$. One expects unitary evolution between the total Hilbert spaces $\HH(\Sigma_u)$. It is clear that
\begin{equation}
    \HH(\Sigma_u) \neq \HH^\rad_U ,
\end{equation}
because the outgoing Hawking modes do not form a complete set of modes on $\Sigma_u$, and $\HH^\rad_U$ is only a subspace of $\HH(\Sigma_u)$. That is, the full Hilbert space consists of more than just the outgoing Hawking modes, even at the end of evaporation. (As a subset of the complete set $\xi^{\eh} \oplus \xi^{+}$, the outgoing Hawking modes ($\xi^{\rad}_{U}$) are missing both the ingoing Hawking modes ($\xi^{\eh}$) and the postevaporation outgoing subset of $\xi^{+}$.) Therefore there is no a priori reason to think the Hawking radiation state $\rho^\rad_U$ should be pure.%
\footnote{Given this failure one might suggest an alternate condition $S_\rad(\infty)=0$ would hold. But this reduces to the question of long term unitarity discussed earlier, as $\xi^+$ are a complete set of modes on $\Sigma_+$.}

Nonetheless, it could still be reasonable to justify the evaporation time unitarity condition based on the time evolution of $S_\rad(u)$. If Page time unitarity were to hold, then so would evaporation time unitarity. Whether this holds is discussed next.

\subsection{Page time unitarity}
\label{sec:page:page}

The ``Page time'' unitarity issue involves the time-dependence of $S_\rad(u)$ in relation to the semiclassical horizon area in the foliation $\Sigma_u$ (Fig.~\ref{fig:foliated}). 

In this foliation $A(u)$, defined as the area of the (outer) apparent horizon on $\Sigma_u$, starts at $A(0) \propto M^2$ and decreases to $A(U)=0$ when evaporation completes.

Page time unitarity will be said to hold if
\begin{equation}
    S_\rad(u) \leq A(u)/4
\end{equation}
at all times.

When Page time unitarity holds, it is usually argued that $S_\rad(u)$ first increases according to Hawking's prediction of thermal emission, until a time (the ``Page time'') when it would surpass $A(u)/4$, after which it decreases according to $S_\rad(u) = A(u)/4$. Then $S_\rad(u)$ is said to follow the ``Page curve''~\cite{Page:2013dx}.

\subsubsection{Argument in favor}

The total Hilbert space $\HH$ consists of ``the black hole plus the Hawking radiation,'' so decomposes as
\begin{equation}
    \HH = \HH^\bh_u \otimes \HH^\rad_u ,
\end{equation}
with reduced densities $\rho^\bh_u$ and $\rho^\rad_u$ in the subsystems. The total system is in a pure state, so
\begin{equation}
    S(\rho^\bh_u) = S(\rho^\rad_u).
\end{equation}
But the thermodynamic (Bekenstein-Hawking) entropy of the black hole is $S_{\rm th}(u) = A(u)/4$. Since thermodynamic entropy is a coarse-grained entropy of the black hole (see \textit{e.g.}~\cite{Page:2013dx,Harlow:2014yka,Polchinski:2016hrw,Almheiri:2020cfm,2019PhRvA..99a0101A,2019PhRvA..99a2103A,2020arXiv200505408S}), it follows that%
\footnote{An alternate justification, that $\log \dim \HH^\bh_u = A(u)/4$, is sometimes assumed to the same effect.}
\begin{equation}
    S(\rho^\bh_u) \leq A(u)/4.
\end{equation}
Thus $S_\rad(u) \leq A(u)/4$.

\subsubsection{The problematic assumption}

The problematic assumption in the preceding argument is the decomposition
\begin{equation}
\label{eqn:problematic}
    \HH = \HH^\bh_u \otimes \HH^\rad_u ,
\end{equation}
where neither $\HH^\bh_u$ nor $\rho^\bh_u$ were given a concrete definition. There are various ways to interpret this statement, depending whether one treats it as a semiclassical or quantum gravitational equation. Each gives a different meaning to ``the entropy of the black hole.''  But none provides a strong justification for 
\begin{equation}
    S_\rad(u) \leq A(u)/4,
\end{equation}
if $S_\rad(u)$ is the von Neumann entropy of semiclassical Hawking modes.

One key point is that the bound $S(\rho^\bh_u) \leq A(u)/4$ derived from coarse-graining is likely to be valid only if $\rho^\bh_u$ represents a full quantum gravitational state---applying this bound in the semiclassical theory requires justifying an identification between ``the Hilbert space of the black hole'' and some space of semiclassical modes.

We analyze the possible interpretations, and their implications, in the following subsections \ref{sec:page:page:interp-1}--\ref{sec:page:page:interp-3}.

\subsubsection{Purely semiclassical interpretation}
\label{sec:page:page:interp-1}

If one works purely within the semiclassical framework, then (\ref{eqn:problematic}) reads
\begin{equation}
\label{eqn:pure-semi-interp}
    \HH(\Sigma_u) = \HH_{\chi_u} \otimes \HH^\rad_u ,
\end{equation}
where $\HH^\bh_u = \HH_{\chi_u}$ is the Hilbert space of ``all the modes except the Hawking modes'' (that is, of $\chi_u$ where $\chi_u \oplus \xi^\rad_u$ is some complete set of modes on $\Sigma_u$).

In this case $\HH^\bh_u$ is the Hilbert space of modes \mbox{$\xi^\eh \oplus \overline{\xi^\rad_u}$}, where $\overline{\xi^\rad_u}$ is the complement in $\xi^+$ of $\xi^\rad_u$ (\textit{cf.}~Figs.~\ref{fig:modes},~\ref{fig:basic-infinity-modes}). This is not the Hilbert space of any relevant partial Cauchy surface, and in particular it is not $\HH(\Sigma^\inn_u)$. Moreover, there is no clear relationship between the modes defining $\HH^\bh_u$ and the horizon area $A(u)$. Indeed, there is no meaningful sense in which the so-called $\HH^\bh_u$ is ``the Hilbert space of the black hole.'' There  is no justification for $S(\rho_{\chi_u}) \leq A(u)/4$, and no reason for Page time unitarity to hold. 

Moreover, Bogoliubov transformations from the modes $\xi^-$ to $\xi^\eh \oplus \xi^+$ in $\MM$ could in principle be directly evaluated, giving a direct calculation of $S_\rad(u)$ under unitary semiclassical evolution. It is unlikely, both in analogy with the standard Hawking calculation, and due to the presence of $\xi^-$ modes straddling the event horizon, that this could lead to a pure state $S_\rad(U)=0$ at the end of evaporation.

\subsubsection{Partially semiclassical interpretation}
\label{sec:page:page:interp-2}

Suppose one interprets $\HH^\rad_u$ as the semiclassical Hilbert space of Hawking modes, but interprets $\HH^\bh_u$ as some quantum gravitational ``full description'' of the black hole (let us denote quantum gravitational Hilbert spaces with a tilde, in this case $\HH^\bh_u = \Tilde{\HH}^\bh_u$). Now there is a fair justification for \mbox{$S(\tilde{\rho}^\bh_u) \leq A(u)/4$}. But another part of the argument breaks down. 

There are two cases, depending whether or not one claims that the quantum gravitational Hilbert space $\Tilde{\HH}^\bh_u$ is equivalent to a semiclassical Hilbert space of modes.

If one does not make such an identification, then there is no guarantee that the Hilbert spaces \mbox{$\Tilde{\HH}^\bh_u \otimes \HH^\rad_u$ and $\HH(\Sigma_u)$} are equivalent. If one starts with a pure state $\ket{\psi}$ in $\HH(\Sigma_u)$, as is typically done, there is no reason for a state in $\HH$ to be pure---if such a state is even defined.

More reasonably, one may claim (perhaps through holography) that the quantum gravitational Hilbert space $\tilde{\HH}^\bh_u$ is equivalent to the Hilbert space of some semiclassical modes $\chi^\bh_u$. These must be part of a complete set $\chi^\bh_u \oplus \chi^{\rm other}_u \oplus \xi^\rad_u$, so that
\begin{equation}
    \HH(\Sigma_u) = \tilde{\HH}^\bh_u \otimes \HH_{\chi_u^{\rm other}} \otimes \HH^\rad_u 
\end{equation}
with  $\tilde{\HH}^\bh_u = \HH_{\chi^\bh_u}$. We have allowed for the presence of some modes $\chi^{\rm other}_u$ that are part of neither the radiation nor the black hole (these could include, for example, outgoing modes after evaporation ends, but we leave them unspecified as they depend on the choice of $\chi^{\bh}_{u}$).

If one assumes these $\chi^{\rm other}_u$ are totally uncorrelated from the rest of the system (schematically, that $\ket{\psi}= \ket{\psi}_{\bh,\rad} \otimes \ket{\psi}_{\rm other}$) then, by adapting the earlier argument, there is a strong justification for $S_\rad(u) \leq A(u)/4$ and Page time unitarity holds.

If one wants to make this claim they should lay out clearly what the modes $\chi^\bh_u$ and $\chi^{\rm other}_u$ are, explain in precisely what sense $\tilde{\HH}^\bh_u = \HH_{\chi_u^\bh}$, and explain why $\chi^{\rm other}_u$ are uncorrelated from the rest of the system.

Lacking a clear and explicit case for this identification, applying the entropy bound $A(u)/4$ to a subset of the semiclassical modes is insufficiently justified.

Nonetheless, the idea that the quantum gravitational Hilbert space $\tilde{\HH}^\bh_u$ can be identified with a space of semiclassical modes is not unreasonable. Later we will return to identifications of this type motivated by holography. In those cases, however, the conclusion  $S_\rad(u) \leq A(u)/4$ still does not necessarily follow. This is because---in the language of this section---either there is no decomposition $\chi^\bh_u \oplus \chi^{\rm other}_u \oplus \xi^\rad_u$ (as would also be the case if one naively identified the quantum gravitational space with the semiclassical \textit{in} region), or because $\chi^{\rm other}_u$ are not uncorrelated from the rest of the system.

\subsubsection{Fully quantum gravitational interpretation}
\label{sec:page:page:interp-3}

If one interprets both pieces of the decomposition as quantum gravitational Hilbert spaces then (\ref{eqn:problematic}) reads
\begin{equation}
    \tilde{\HH} = \tilde{\HH}^\bh_u \otimes \tilde{\HH}^\rad_u .
\end{equation}
This is the case when the black hole Hilbert space is described through AdS/CFT.

In this case there is justification for $S(\tilde{\rho}^\rad_u) \leq A(u)/4$. However, now it is not clear that the quantum gravitational space $\tilde{\HH}^\rad_u$ relates to the space of semiclassical Hawking modes $\HH^\rad_u$. In other words, with this interpretation, $S(\tilde{\rho}^\rad_u) \neq S_\rad(u)$. 

So while the Page curve likely arises in quantum gravity, that may not imply the same for the semiclassical modes. Understanding that correspondence requires further investigating the relationship of the semiclassical and quantum gravitational Hilbert spaces, which will be considered in Sec.~\ref{sec:holographic}.

\subsubsection{Summary}

Naively decomposing $\HH(\Sigma_u) = \HH^\bh_u \otimes \HH^\rad_u$ leads to the conclusion $S_\rad(u) \leq A(u)/4$, implying that $S_\rad(u)$ follows a Page curve. But closer inspection reveals that if $S_\rad(u)$ is meant to be the von Neumann entropy of semiclassical Hawking modes, then in any interpretation either the decomposition itself is invalid, or the conclusion does not follow from it. Thus the claim that $S_\rad(u)$ follows a Page curve is weak, and other curves for the semiclassical $S_\rad(u)$, including the traditional Hawking curve, may be consistent with unitarity. On the other hand, entropies describing quantum gravitational degrees of freedom may still follow a Page curve, as discussed later.

\section{Entropy of the ``In/Out'' regions}
\label{sec:in-out}

In the previous section, questions of unitarity were framed in terms of $S_\rad(u)$, the von Neumann entropy of Hawking modes at future infinity. That entropy is distinct from, though sometimes conflated with, the von Neumann entropy
\begin{equation}
    S_\out(u) \equiv S(\rho^\out_u)
\end{equation}
of fields in the \textit{out} region. Here $\rho^\out_u$ is the reduced density matrix on $\Sigma^\out_u$ (Fig.~\ref{fig:foliated}). Each is defined in terms of a different mode decomposition: $S_\rad(u)$ in terms of $\xi^\eh \oplus \xi^+$ (Fig.~\ref{fig:modes}), and $S_\out(u)$ in terms of $\eta^\inn_u \oplus \eta^\out_u$ (Fig.~\ref{fig:in-out-modes}).

\begin{figure}
    \centering
    \includegraphics[height=2.5in]{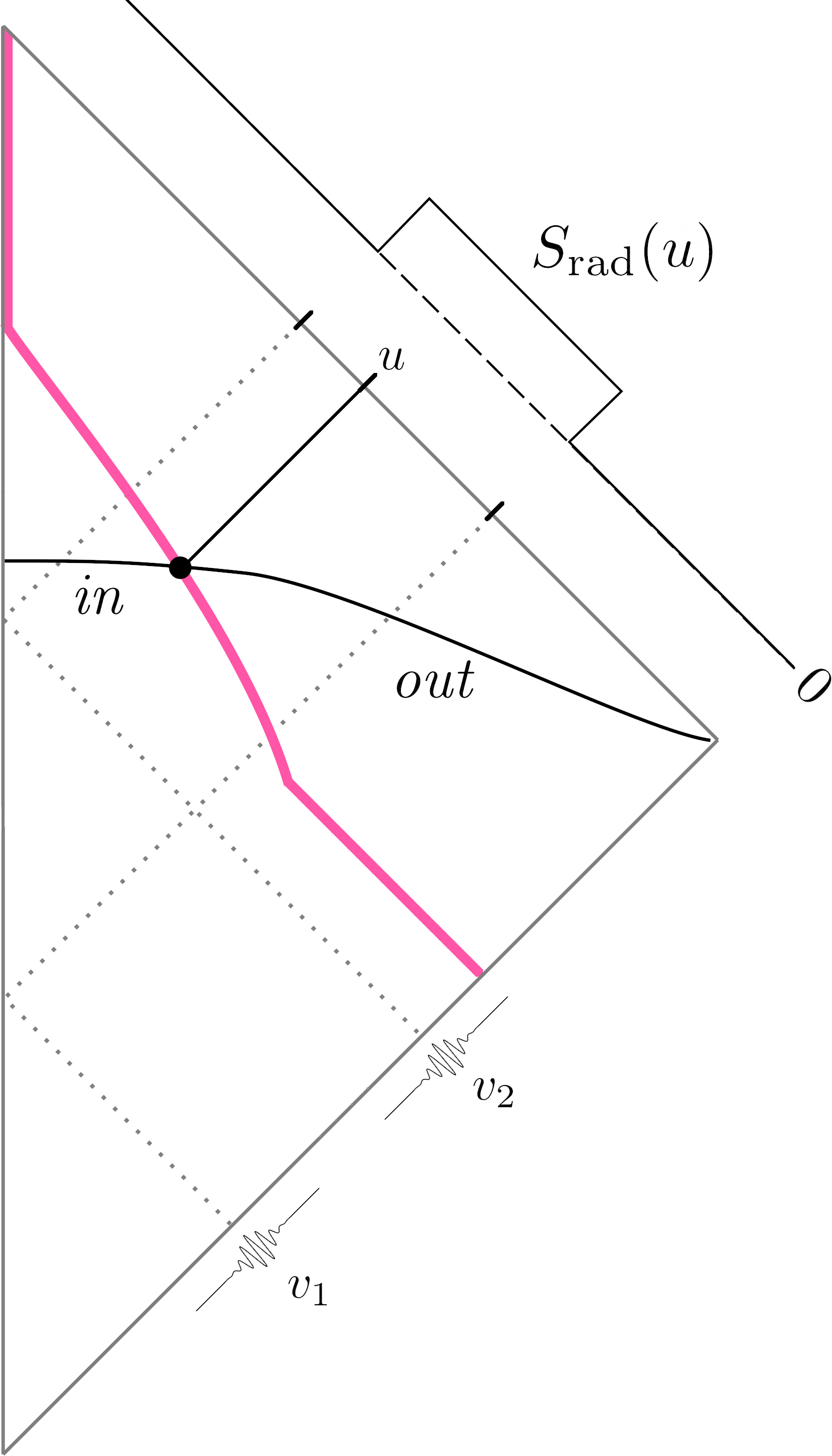}
    \caption{Minkowski space with fields in a state $\ket{\psi}=\ket{1_{v_1} 1_{v_2}}$ in terms of wavepacket modes similar to $\xi^-$. $S_\rad(u)$ rises when an entangled mode reaches infinity, then is purified to zero when its partner arrives. $S_\out(u)$, the von Neumann entropy in the \textit{out} region (whose boundary is chosen to be analogous to $\Sext$ in $\MM$), is UV-divergent due to vacuum entanglement, but can be argued to be related to $S_\rad(u)$ after renormalizing.}
    \label{fig:minkowski-example}
\end{figure}

These entropies, $S_\rad(u)$ and $S_\out(u)$, have vastly different character, as can be seen in the simple Minkowski space example of Fig.~\ref{fig:minkowski-example}. In that example, $S_\rad(u)$ begins at zero, increases as entangled modes arrive at infinity, then is purified back to zero by the later radiation. Meanwhile, $S_\out(u)$ is infinite at all times, with a UV-divergent leading order (``vacuum'') contribution proportional to the area of its boundary~\cite{Srednicki:1993im,Holzhey:1994we,Calabrese:2004eu}.

Despite this basic difference, these entropies may be related. There exist in the literature a number of plausible arguments~\cite{Holzhey:1994we,Page:2013dx,Almheiri:2020cfm} (see also~\cite{Casini:2008cr}) that
\begin{equation}
    S_{\out}(u) \approx S_{\rm vac}(u) + S_\rad(u),
\end{equation}
or equivalently, that $S_\out^{\rm ren}(u) \approx S_\rad(u)$ after renormalizing by subtracting out the vacuum term.%
\footnote{If one displaces the \textit{in/out} boundary surface outward to a line of constant radius outside the horizon, as done for instance in~\cite{Almheiri:2020cfm}, this renormalization should amount to subtracting a divergent constant (proportional to the constant area of the boundary).}

\begin{figure}[t]
    \centering
    \includegraphics[height=2.3in]{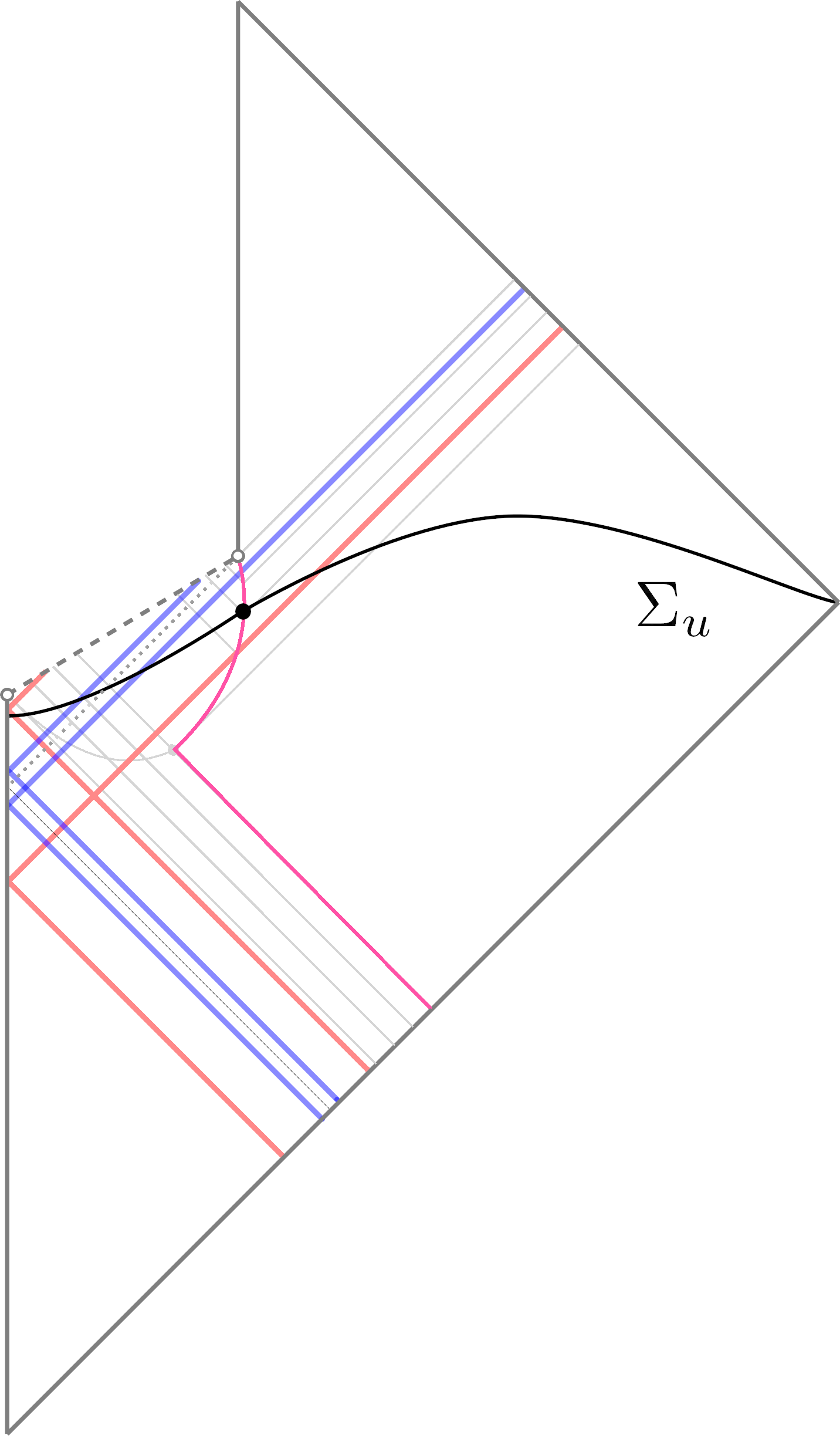}
    \caption{Early (red) and late (blue) pairs of entangled Hawking modes. As $u \to U$, later outgoing Hawking partners emerge into the \textit{out} region, similar to the Minkowski space example of Fig.~\ref{fig:minkowski-example}.}
    \label{fig:hawking-pairs}
\end{figure}

Such arguments are generally based on the idea that, if one partner in a pair of Hawking modes has significant support only in the \textit{out} region, that mode contributes its entropy to the \textit{out} region. As $u \to U$ more outgoing Hawking partners emerge into the \textit{out} region causing an evolution of $S_\out(u)$. This scenario is depicted in Fig.~\ref{fig:hawking-pairs}. One can make an analogous argument in the Minkowski space example of Fig.~\ref{fig:minkowski-example}.%
\footnote{In that example the argument would suggest that $S_{\out}(u) = S_{\rm vac}(u) +1$ bits for $u_1 < u < u_2$, and $S_{\out}(u) = S_{\rm vac}(u)$ otherwise. Note that this cannot be exactly true as the entangled modes each have finite width.}

Now we return to the question of Page/evaporation time unitarity, and its relation to the firewall problem, this time in the context of $S_\out(u)$.

\subsection{Page time unitarity again}

Suppose one identifies the quantum gravitational black hole Hilbert space with the semiclassical Hilbert space of modes behind the horizon, $\tilde{\HH}^\bh_u = \HH(\Sigma^\inn_u)$. Then, since von Neumann entropy is not greater than thermodynamic (coarse-grained) entropy, one would expect $S(\rho^\inn_u) \leq A(u)/4$. Since the total bipartite state is pure, this also implies $S_\out(u) \leq A(u)/4$, suggesting that $S_\out(u)$ would follow a Page curve. If this were true it would imply a product state $\ket{\psi}_\inn \otimes \ket{\psi}_\out$ at the end of evaporation, and thus a firewall at the late time apparent horizon.

As with the discussions in Sec.~\ref{sec:page-time}, the identification $\tilde{\HH}^\bh_u = \HH(\Sigma^\inn_u)$ must be called into question, and requires a more concrete justification. Naively applying holographic principles may suggest this identification, but more detailed studies based on entanglement wedge reconstruction suggest a different one (see Sec.~\ref{sec:holographic}).

The bound $S(\rho^\inn_u) \leq A(u)/4$ also cannot be directly justified through a Bousso bound~\cite{Bousso:1999xy} (or Bekenstein bound~\cite{Bekenstein:1980jp}), since the only converging lightsheet from a point on the apparent horizon terminates at the spacelike singularity. This was pointed out earlier by Rovelli~\cite{Rovelli:2019tbl}.

Moreover, $S(\rho^\inn_u)$ is formally infinite, so the thermodynamic bound must be applied either after a UV cutoff, or after renormalizing by subtracting the vacuum term. In the first case the bound seems to be a statement mainly about the dominant vacuum term, and not about the entropy of Hawking radiation. Moreover, the fact that the Page curve begins at zero seems to preclude it from including the vacuum entropy. In the second case, it is not clear why the coarse-grained bound on von Neumann entropy should still be relevant after renormalization.

In this light, evaporation time and Page time unitarity can be expected for neither $S_\out(u)$ nor $S_\rad(u)$, barring some improved justification.

\begin{figure}[t]
    \centering
    \includegraphics[height=2.3in]{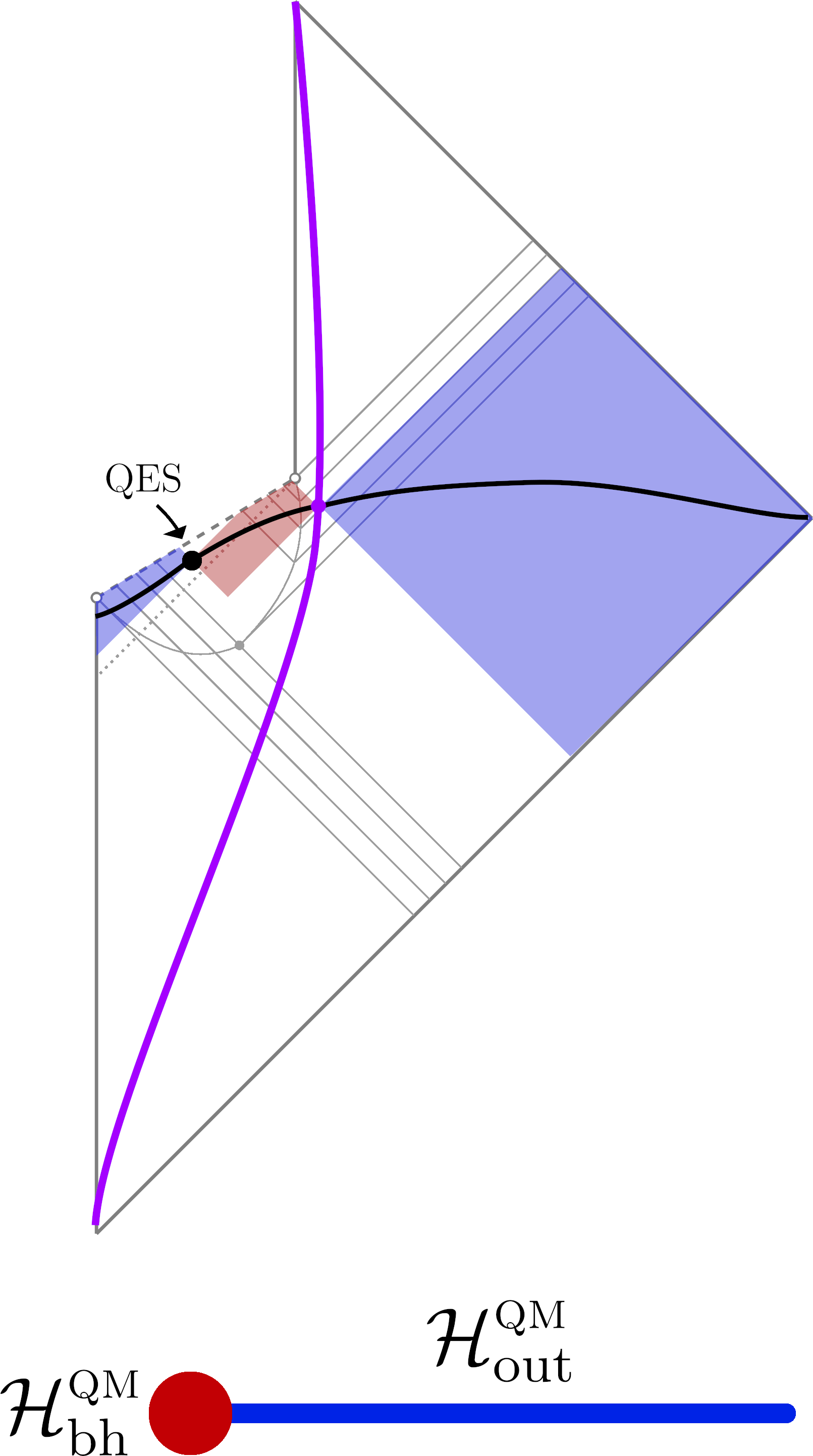}
    \caption{A ``boundary'' theory on $\HH^\qm = \HH^\qm_\bh \otimes \HH^\qm_\out$ with a semiclassical ``bulk'' dual. Depicted is a time after the Page time during evaporation. The entanglement wedge of $\HH^\qm_\bh$ (red fill) is bounded by a cutoff surface (purple line) and quantum extremal surface (QES). The entanglement wedge of $\HH^\qm_\out$ (blue fill) includes an exterior region and an ``island'' near $r=0$. Below is an illustration of the boundary theory where the holographic black hole Hilbert space (red dot) is coupled to a CFT (blue line). See Fig.~18 of \cite{Almheiri:2020cfm} and Fig.~2 of \cite{Almheiri:2019hni}, on which this picture is based, for details.}
    \label{fig:wedge}
\end{figure}

\section{Connection to holographic quantum gravity}
\label{sec:holographic}

Throughout earlier sections, the question was raised of what correspondence exists between semiclassical Hilbert spaces and underlying quantum gravitational ones. Recent holographic studies~\cite{Penington:2019npb,Almheiri:2019hni,Almheiri:2020cfm,Almheiri:2019psf} (based on the AdS/CFT correspondence~\cite{Maldacena:1997re}) suggest a solution. 

Given a quantum ``boundary'' theory on \mbox{$\HH^\qm = \HH^\qm_\bh \otimes \HH^\qm_\out$} whose semiclassical ``bulk'' dual is a forming and evaporating black hole, the boundary Hilbert spaces $\HH^\qm_\bh$ and $\HH^\qm_\out$ each determine an entanglement wedge in the bulk. These entanglement wedges are illustrated (based on the calculations of~\cite{Almheiri:2019hni,Almheiri:2020cfm}), during evaporation after the Page time, in Fig.~\ref{fig:wedge}.

In this context one may calculate boundary von Neumann entropies $S(\rho^\qm_\bh)$ and $S(\rho^\qm_\out)$, and show that they are equal and follow a Page curve~\cite{Penington:2019npb,Almheiri:2019hni,Almheiri:2020cfm}. These boundary entropies are related to bulk entropy in their entanglement wedge through a quantum extremal surface (quantum Ryu-Takayanagi) prescription~\cite{Ryu:2006bv,Hubeny:2007xt,Engelhardt:2014gca,Dong:2016eik}. In particular,
\begin{equation}
    S(\rho^\qm_\out) = S_{\rm bulk}(\Sigma^\qm_\out) + \tfrac{A_{\textsc{qes}}}{4},
\end{equation}
where $S_{\rm bulk}(\Sigma^\qm_\out)$ is the von Neumann entropy of bulk fields on any surface $\Sigma^\qm_\out$ that is a Cauchy surface for the entanglement wedge of $\HH^\qm_\out$. (That is, on a Cauchy surface for just the blue region in Fig.~\ref{fig:wedge}, including both the exterior region and island. The intersection of the black line with both parts of the blue region is one such surface.) $A_{\textsc{qes}}$ is the area (in Planck units) of the appropriate quantum extremal surface.

Thus the boundary entropies, which follow the Page curve, do dictate the semiclassical entropy in certain regions---but these regions are not the ones usually naively identified as \textit{in} and \textit{out}. That is, $S(\rho^\qm_\out)$ does follow the Page curve, but the entropy $S_\out(u)$ in the bulk \textit{out} region need not.%
\footnote{The \textit{out} region in the holographic calculations is defined by a fixed radius surface outside all horizons, rather than by our $\Sext$, but the conclusion is the same.}

In particular, these studies suggest that, after the Page time, the bulk fields in each entanglement wedge (Fig.~\ref{fig:wedge}) have negligible von Neumann entropy (after subtracting a vacuum term) to leading order. This is consistent with a semiclassical Hawking curve in the bulk \textit{out} region, arising from entanglement between the \textit{out} region and the island.

Moreover, as depicted in Fig.~18 of~\cite{Almheiri:2020cfm}, after evaporation ends the entanglement wedge of $\HH^\qm_\out$  contains the region behind the event horizon. Thus the semiclassical state on a late spatial slice like $\Sigma_+$ (Fig.~\ref{fig:long-term}) need not be pure, even though $\HH^\qm_\out$ is in a pure state.

This suggests that semiclassical long term, evaporation time, and Page time unitarity may all fail, even when a Page curve arises in an underlying unitary theory of quantum gravity.%
\footnote{Some other studies (\textit{e.g.}~\cite{Akers:2019nfi}) have suggested that the boundary Page curve and bulk Hawking curve are contradictory. This relies on identifying the bulk and boundary entropies in a way that does not follow from entanglement wedge reconstruction. This also assumes that a bulk Hawking curve in $\MM$ fundamentally violates unitarity, which we have argued against.}%
$^,$%
\footnote{Recently it has also been argued \cite{Marolf:2020rpm}, based on the path-integral quantum gravity of an ensemble of black holes, that a version of Page time unitarity arises effectively within superselection sectors of the theory. In that approach the notion of quantum gravitational unitarity for an individual black hole may differ somewhat from the above discussions.}

Given these two levels of description, what will be measured by an observer at infinity? This depends on precisely what is meant by ``observer at infinity,'' in particular whether such an observer interacts locally with bulk or boundary operators. Clearly an observer with access to all boundary observables can deduce all information about the state.%
\footnote{As with any quantum system, an observer with access to these observables would still need to reconstruct the state through tomography on an ensemble in order to gain full information about the state.}
However, if one conceives of an observer at infinity as one that observes itself outside a spatially distant gravitating object, it seems implicit that such an observer is interacting with bulk observables. In contrast, it is not clear in precisely what sense an observer in the boundary theory can be described as being outside a spatially distant gravitating object, given the nonlocal boundary encoding of interior and exterior bulk degrees of freedom (see \textit{e.g.}~\cite{Chen:2019uhq}). Further clarifying what bulk or boundary observables might be realistically measured in experiments (\textit{i.e.}~which type of operators can ``we'' measure) is a useful topic of continued study.

\section{Conclusions}

The correct statement of the principle of unitarity depends at what level a theory is described. In semiclassical gravity, it demands a unitary evolution between states in $\HH(\Sigma_u)$, the Hilbert space of quantum fields on a series of Cauchy surfaces. In quantum gravity, it demands unitary evolution of states in $\HH^{\qm}$, an underlying quantum mechanical Hilbert space from which spacetime and gravity may emerge.

We have argued that even if unitarity holds in both senses described above, the more commonly invoked notions of long term, evaporation time, and Page time ``unitarity'' may all be violated. In other words, neither ``information loss at infinity'' nor a semiclassical ``Hawking curve'' necessarily signify unitarity violation.%
\footnote{As these are the forms of ``unitarity'' usually assumed in the argument for firewalls, this implies there is no need for firewalls to form.}

One key aspect of the argument was the distinction between semiclassical and quantum gravitational degrees of freedom---holographic calculations suggest that a Page curve is present at the quantum gravitational level, but not necessarily at the semiclassical level.

We see four ways to refute our conclusions about unitarity. One could claim that: (1) No semiclassical theory accurately describes black hole formation and evaporation; (2) There is a useful semiclassical theory but $\MM$ is a poor approximation of it; (3) The semiclassical framework above contains faulty assumptions or unjustified steps; (4) Within the above framework, there is a stronger justification for long term, evaporation time, or Page time unitarity that was not considered. It would be useful to distinguish between these possibilities in claims that these forms of unitarity are restored.

On occasion other entropies are studied besides the ones considered here. In that context, one might introduce some entropy related to black hole evaporation, find that it deviates from a Page curve, and claim that this signifies an information problem. Then, one can introduce some other (perhaps very different) quantity, also called ``entropy,'' which does follow a Page curve, thereby resolving the problem. Generalizing the present work, we suggest that an entropy deviating from a Page curve is not necessarily problematic, and any unitarity problem that arises should be made clear and explicit. Further, any entropies introduced in these analyses should be carefully related to a particular meaning of the black hole Hilbert space.

Here we studied the problem in a spacetime $\MM$ with singularity. A number of other papers have argued for nonsingular models (like Fig.~\ref{fig:regular}), where quantum effects regulate the singularity. The common objection to such models is the claim: \textit{A unitarity problem arises at the Page time, when singular and nonsingular models are equivalent.} If that were true, regularized models would be irrelevant to the information problem.

Our conclusions amount to an argument against this objection, affirming the viability of regular models. Similar arguments were made recently by Ashtekar~\cite{Ashtekar:2020ifw} using a regular model inspired by loop quantum gravity that coincides with $\MM$ in the semiclassical region (our $D(\Sigma_-)$). In that paper another form of the Page time argument based on ``energy budget per mode'' was also refuted. In regular spacetimes one expects long term unitarity to be restored, while evaporation time and Page time unitarity remain violated.

Ultimately there is no guarantee that \textit{any} semiclassical spacetime can fully represent the black hole evaporation process. Nonetheless, use of spacetimes like $\MM$ is prevalent in the literature.

We emphasize that even if one does believe $\MM$ is a useful evaporation model, black hole evaporation is not paradoxical. There is no fundamental contradiction between unitarity and relativity. A contradiction only arises if one considers limited forms of semiclassical unitarity that, on closer inspection, are poorly motivated. 

On the other hand, the fact that long term unitarity is given up in $\MM$ is a sign of its pathologies (lack of global hyperbolicity and the pathology of future null infinity). It does seem reasonable to hope evaporation will be described by a semiclassical theory with a scattering matrix from past to future infinity (unless there arise significant correlations between matter and geometry, or matter and sub-Planckian degrees of freedom). But such a theory will not include something like $\MM$ as a background. 

\vspace{1mm}

\begin{acknowledgments}
This research was supported by the Foundational Questions Institute (FQXi.org), of which AA is Associate Director, and by the Faggin Presidential Chair Fund.
\end{acknowledgments}



\bibliography{inspire,ads}






\end{document}